\documentclass[12pt]{article}
\usepackage[margin=0.95 in]{geometry}
\usepackage{amsmath} 
\usepackage{amssymb,amsfonts}
\usepackage[all]{xy}
\usepackage{graphicx}
\usepackage{amsmath}
\usepackage{amssymb}
\usepackage{float}
\usepackage{array}
\usepackage{tikz}
\usepackage{mathtools}
\usepackage{mathrsfs} 
\usepackage[hidelinks]{hyperref}
\usepackage{cite}
\numberwithin{equation}{section}
\usepackage[utf8x]{inputenc}

\newcommand{\be}{\begin{equation}}
\newcommand{\ee}{\end{equation}}
\def\bea{\begin{eqnarray}}
\def\eea{\end{eqnarray}}

\numberwithin{equation}{section}
\numberwithin{table}{section}

\begin{document}

\hypersetup{pageanchor=false}
\begin{titlepage}
\vbox{
    \halign{#\hfil         \cr
           } 
      }  
\vspace*{9mm}
\begin{center}
{\Large \bf  Mixed Solutions to the  Liouville Equation
 }

\vspace*{4mm}

{\em with  Applications to 
The Gravitational 
 Two Body Problem in Three Dimensions
}
\vspace*{15mm}

{\large  Sujay K. Ashok$^{a,b}$ and Jan Troost$^c$}
\vspace*{8mm}

$^a$The Institute of Mathematical Sciences, \\
		 IV Cross Road, C.I.T. Campus, \\
	 Taramani, Chennai, India 600113

\vspace{.6cm}

$^b$Homi Bhabha National Institute,\\ 
Training School Complex, Anushakti Nagar, \\
Mumbai, India 400094

\vspace{.6cm}

$^c$Laboratoire de Physique de l'\'Ecole Normale Sup\'erieure \\ 
 \hskip -.05cm
 ENS, CNRS, Universit\'e PSL,  Sorbonne Universit\'e, \\
 Universit\'e  Paris Cit\'e  \\
 24 rue Lhomond,
 F-75005 Paris, France	 

\vspace*{0.8cm}
\end{center}

\begin{abstract}
We enlarge the set of explicit classical solutions to the Liouville equation with three singularities to the cases with mixed hyperbolic and elliptic monodromies. We analyze the large hyperbolic monodromy limit of the solutions and  the farthest geodesics looping one hyperbolic singularity. These two-dimensional geometries describe a time-symmetric spatial slice of a solution to three-dimensional general relativity. The geodesics are reinterpreted as snapshots of horizons of evolving black holes. We study the spatial slice with three horizons of very heavy black holes in some detail. 
We  use uniform saddle point integration  to present the Liouville and  heavy black hole geometries in terms of simpler special functions. These make a detailed analysis of mixed particle and black hole geometries possible.

\end{abstract}

\end{titlepage}
\hypersetup{pageanchor=true}

\tableofcontents

\section{Introduction}
The two-dimensional Liouville equation 
\begin{equation}
\partial \bar{\partial} \phi  = \frac{e^{\phi}}{2} 
\label{IntroLiouville}
\end{equation}
 for a real function $\phi(z,\bar{z})$ of two complex conjugate variables 
 was solved in considerable generality by Joseph Liouville in 1853 \cite{Liouville}. The Liouville equation has applications to a multitude of fields in mathematics and physics. The classical solutions to the equation received much attention -- see for example \cite{Bilal:1987cq,Seiberg:1990eb,Zamolodchikov:1995aa}. Nevertheless, a lot of properties of the classical solutions still deserve to be  understood better and we aim to contribute to this basic aim in mathematical physics. Our insights will  ripple through to, among others, the application of the Liouville equation to three-dimensional gravity.

 In this paper, we  consider solutions to the Liouville equation with three point-like classical sources either specified by delta-function source contributions or by boundary conditions near the insertion. 
 In section \ref{ClassicalSolutions}, we study solutions of the Liouville equation (\ref{IntroLiouville}) with three hyperbolic,  elliptic or mixed type insertions. We match the all-hyperbolic and all-elliptic solutions in the literature and clarify how they generalize to mixed configurations. We provide  intuitive insight into the resulting two-dimensional geometries.  In all cases, we study  limits of the solutions in which  the hyperbolic sources become large in section \ref{LimitSolutions}. A uniform saddle point approximation to hypergeometric functions is employed to derive various explicit metrics in terms of simpler functions. We moreover analyze the behavior of the  geodesics that loop a single hyperbolic singularity. 
 Starting from section \ref{3dGravity}, we discuss the application of the classical solutions to the Liouville equation to three-dimensional gravity with a zero or negative cosmological constant. Since this is but one application of the Liouville equation, we postpone an introduction to this specific field to section \ref{3dGravity}. In a nutshell, the two-dimensional geometries described by the Liouville field describe a time-symmetric slice of a three-dimensional geometry with black hole singularities corresponding to hyperbolic sources and point particle singularities to elliptic sources. 
We conclude in section \ref{Conclusions} and %
collect some technical details regarding the classical Liouville solutions in an Appendix.

\section{Classical Solutions to the Liouville Equation}
\label{ClassicalSolutions}
\label{Liouville}

Classical solutions to the Liouville equation have applications in many  domains including two-dimensional conformal field theory, Euclidean geometry and string theory. They are  worth describing in detail. 
The classical solution with three hyperbolic monodromies was described with pedagogical clarity in \cite{Hadasz:2003kp,Hadasz:2003he} and \cite{Firat:2021ukc} while for the case of three elliptic monodromies, the solution was succinctly given in \cite{Welling:1997fw}. We review how to obtain these solutions and add a detailed description of the solutions with three points of monodromy for the cases in which the monodromies are of mixed hyperbolic and elliptic type.\footnote{
Cases that involve parabolic monodromies are also interesting. } The solutions of mixed type have worthwhile applications in conformal field theory and two- and three-dimensional geometry. We concentrate on the latter field of application from section \ref{3dGravity} onward.

\subsection{Doublets with Group Monodromies}
Another standard form for the Liouville equation is
\begin{equation}
 \partial \bar{\partial} \Phi + \sqrt{2} b M e^{\sqrt{2} b\Phi}  =0 \, . \label{LiouvilleIntro} \, 
 \end{equation}
 By shifting the field $\Phi$ we can rescale the parameter $M$ by a positive number and by rescaling $\Phi$ we can  eliminate the parameter $b$. The latter parameter however is also present in implicit boundary conditions or in a standard modification of the equation on a curved surface and is a physical parameter of the corresponding quantum conformal field theory \cite{Bilal:1987cq,Seiberg:1990eb,Zamolodchikov:1995aa}. 
 
 We can modify the Liouville equation by adding source 
 terms on the right hand side of equation (\ref{LiouvilleIntro}):
 \begin{equation}
 \partial \bar{\partial} \Phi + \sqrt{2} b M e^{\sqrt{2} b\Phi}  =  \sum_{i=1}^3 \beta_i \delta^{(2)} (z-z_i)  \, .
 \end{equation}
 For ease of calculation, we do simplify the Liouville equation by some rescalings.  We define the field ${\phi} = \sqrt{2} b \, \Phi$ to find:
\begin{equation}
 (\sqrt{2} b)^{-1} \partial \bar{\partial} {\phi} + \sqrt{2} b M e^{{\phi}}  =  \sum_{i=1}^3 \beta_i \delta^{(2)} (z-z_i)  \, .
 \end{equation}
Next, we pick the parameter $M$ such that the relative coefficient between the two terms on the left hand side is $-1/2$:
 \begin{equation}
 M = -\frac{1}{4 b^2} 
\end{equation} and write:
\begin{equation}
  \partial \bar{\partial} {\phi} - \frac{1}{2} e^{{\phi}}  =  \sqrt{2} b \sum_{i=1}^3 \beta_i \delta^{(2)} (z-z_i)  \, . \label{StandardLiouville}
 \end{equation}
We will consider parameters  $  \gamma_i = (\sqrt{2} b) \beta_i$  that are  finite. When the coefficients $\gamma_i$ are real and within a given range,  the sources are elliptic. When we continue the parameters to imaginary values, we have a hyperbolic sources. The latter are better described in terms of boundary conditions. 
 To describe the solutions with sources, we will study singular solutions to the  sourceless Liouville equation  
\begin{equation}
\label{Liouvilleeqn}
 \partial \bar{\partial} \phi = \frac{1}{2} e^{\phi} \, .
 \end{equation}
 The field $\phi$ can be constructed in terms of an auxiliary doublet of fields $\psi^\pm$ that have monodromies around points while the Liouville field $\phi$ remains uni-valued.  In this section, we begin with the case in which all these monodromies are hyperbolic  \cite{Hadasz:2003he,Firat:2021ukc} and then discuss the elliptic and mixed cases as well.

The key point here is to relate the problem of finding the classical solutions to the non-linear Liouville equation (with prescribed boundary conditions) to an auxiliary linear differential equation. This has been well studied, both in the physics setting of quantum Liouville theory \cite{Bilal:1987cq,Seiberg:1990eb,Zamolodchikov:1995aa} and also in the mathematical setting of the uniformization problem. We  only emphasize the points most relevant to us. The auxiliary equation is a holomorphic differential equation, of the form:
\be 
\partial_z^2\psi(z)  + \frac{1}{2} T_{\phi}(z)\psi(z) = 0~, 
\ee
where $T_{\phi}$ is the (semi-classical) stress tensor associated to Liouville theory:
\be 
T_{\phi}(z) = -\frac12(\partial\phi)^2 + \partial^2\phi~. 
\ee 
Using the Liouville equation in \eqref{Liouvilleeqn}, it is simple to check that $\psi(z) = e^{-\frac{\phi(z)}{2}}$ satisfies the linear differential equation. This is a real and single valued solution of the differential equation. However the most general linearly independent, normalized solutions $\psi^{\pm}(z)$ of the linear differential equation are complex and multi-valued. We shall  write the function $e^{-\frac{\phi(z)}{2}}$ in terms of the multi-valued solutions $\psi^{\pm}(z)$. In what follows we shall assume that the multi-valuedness is of a particular type:   the monodromy around  singularities belongs to the group SL$(2, \mathbb{R})$ or, equivalently, to the group SU$(1,1)$. The two independent solutions to the second order linear differential equation will be either an SL$(2,\mathbb{R})$ doublet or an SU$(1,1)$ doublet. The associated formalisms are equivalent but either one or the other is slightly handier in calculations with monodromies of either hyperbolic or elliptic type. 

\subsubsection{The Auxiliary $SL(2,\mathbb{R})$  Doublet}
The expression for the Liouville field $\phi$ in terms of the doublet $\psi^\pm$ can take two equivalent but slightly differing forms.
Firstly, if the fields $\psi^\pm$ are taken to form an $SL(2,\mathbb{R})$ doublet then the expressions are as follows. Suppose we have 
 linearly independent and normalized solutions $\psi^\pm$ to the Fuchsian equation. The $SL(2,\mathbb{R})$ doublet $\psi^\pm$ is multi-valued around regular singularities.  The field $e^{-\frac{\phi}{2}}$ is single valued and equal to 
 \begin{equation}
 e^{-\frac{\phi}{2}} = \pm  \frac{i}{2} (\overline{\psi^-} \psi^+ -\overline{\psi^+} \psi^-) \, . 
\end{equation}
The Liouville field is invariant under $SL(2,\mathbb{R})$ transformations because the fields $\psi^\pm$ transform as a doublet.
We also have an associated two-dimensional metric that arises from viewing the Liouville field as a conformal factor:
\begin{equation}
ds^2 =  e^{\phi} dz d\bar{z} = - \frac{4 \, dz d\bar{z}}{(\overline{\psi^-} \psi^+ -\overline{\psi^+} \psi^-)^2} \, . 
\end{equation}
The normalizing Wronskian condition on the doublet of solutions reads:
\begin{equation}
\partial \psi^+ \psi^- - \partial \psi^- \psi^+ = 1 \, .
\end{equation}
Both the Liouville field and therefore also the metric 
are invariant if the basis vector $\psi^\pm$ undergoes an $SL(2,\mathbb{R})$ monodromy as it is transported around a given excised singular point.

\subsubsection{The Auxiliary $SU(1,1)$ Doublet}

 We can also write the solution in terms of a $SU(1,1)$ doublet $\xi^\pm$. 
The $SU(1,1)$ doublet  $\xi^\pm$ and the $SL(2,\mathbb{R})$ doublet $\psi^\pm$ are related through the change of basis  $\xi=C \psi$ where $C$ is the Cayley matrix
\begin{equation} 
C = \frac{1}{\sqrt{2}}\begin{pmatrix}
    1 & i \\
    i & 1
\end{pmatrix} \, .
\end{equation}
An $SL(2,\mathbb{R})$ monodromy matrix $M$ is equivalent to a $SU(1,1)$ monodromy matrix $N$ through the equality $M=C^{-1} N C$.
In this formalism, the metric reads:
\begin{equation}
e^\phi = \frac{4 |\partial f|^2}{(1-|f|^2)^2} = \frac{4}{(|\xi^+|^2 - |\xi^-|^2)^2} \, ,
\end{equation}
where the function $f$ 
\begin{equation}
f = \frac{\xi^+}{\xi^-}
\end{equation}
is a ratio of $SU(1,1)$ doublet components $\xi^\pm$. 
The calculation of the metric  exploits  the Wronskian condition:
\begin{equation}
\xi^+ \partial \xi^--\xi^- \partial \xi^+ = 1  \, .
\end{equation}
The Wronskian condition does not fix a complex factor $e^{\pm \frac{iv}{2}}$ that we can multiply into a doublet $\psi^\pm$ or $\xi^\pm$. The
$SU(1,1)$ transformations that  leave the metric invariant are
\begin{equation}
f  \rightarrow \frac{\beta f + \gamma}{\bar{\gamma} f+ \bar{\beta}}
\end{equation}
where the transformation matrix
\begin{equation}
N = \begin{pmatrix}
    \beta & \gamma \\
    \bar{\gamma} &  \bar{\beta}
\end{pmatrix} \, 
\end{equation}
is an element of the group $SU(1,1)$. 
Whether we use the $SL(2,\mathbb{R})$ or $SU(1,1)$ formalism to describe the Liouville solutions is immaterial. Still, in practice, one may prefer either one or the other to express the local properties of a solution.

\subsubsection{Our Basis Picks}

For hyperbolic singularities the monodromy can be diagonalized in the $SL(2,\mathbb{R})$ formalism and it is technically slightly easier to work in this formalism near one of these singularities. Near a hyperbolic singularity (at $z=0$) we will work with a basis of solutions that can be expanded locally as:
\begin{equation}
\psi^\pm = \frac{e^{\pm \frac{iv}{2}}}{\sqrt{i \lambda}} z^{\frac{1 \pm i \lambda}{2}} (1+O(z))
\end{equation}
with $\lambda$ real. 
The function $\rho$
\begin{equation}
\label{fdefnz0}
\rho(z) = \left(\frac{\psi^+}{\psi^-}\right)^{\frac{1}{i \lambda}} \, 
\end{equation}
is then single valued around the singularity $z=0$. 
The Liouville field is also single valued and can be written in terms of the function $\rho$ as 
\begin{equation}
e^\phi =  \frac{|\partial \lambda \log \rho|^2}{\sin^2 \lambda \log |\rho(z)|} \, . \label{MetricNearHyperbolic}
\end{equation}
By contrast, we prefer to work in the $SU(1,1)$ formalism near an elliptic singularity since we can then once more diagonalize the monodromy by picking a basis that reads
\begin{equation}
\xi^\pm = \frac{e^{\pm  \frac{i {v}}{2}}}{\sqrt{\alpha}} z^{\frac{1 \pm \alpha}{2}} (1+O(z))
\end{equation}
near the singularity (here at $z=0$), with a  parameter $\alpha$ that is real. We introduce another function $\rho$ 
\begin{equation}
\rho(z) = \left(\frac{\xi^+}{\xi^-}\right)^{\frac{1}{\alpha}} \, 
\end{equation}
which is single valued near an elliptic singularity. 
The metric in terms of the function $\rho$ now reads:
\begin{equation}
ds^2 = \frac{\alpha^2 \partial \rho \bar{\partial} \bar{\rho}}{|\rho|^2 \sinh^2 (\alpha \log |\rho|)} 
\, . \label{MetricNearElliptic}
\end{equation}
While the $SL(2,\mathbb{R})$ and $SU(1,1)$ formalisms are  related, as are some properties of the solutions near points of elliptic or hyperbolic monodromy,  we have paid careful attention in writing out the details of the two cases since some crucial geometric and physical properties of hyperbolic and elliptic singularities are  distinct.

\subsection{The Four Classes of Combinations of Three Monodromies}
We will study solutions to Liouville's equation with three singularities. 
We can have zero, one, two or three points around which the auxiliary doublet has hyperbolic monodromies and we will assume that the other points exhibit elliptic monodromies. Thus, we have four classes of solutions:  all-hyperbolic, hyperbolic-elliptic-hyperbolic, elliptic-elliptic-hyperbolic and all-elliptic. We treat each of these cases in turn.

\subsubsection{Three Hyperbolic Monodromies}
\label{AllHyperbolicSubsection}
We briefly review the case where we have three hyperbolic monodromies for the $SL(2,\mathbb{R})$ doublet around  singular points chosen to lie at $z=0,1$ and $\infty$. This case was described in pedagogical detail in the references \cite{Hadasz:2003he, Firat:2021ukc}. We label the three points with indices $i=1,2,3$. 
The second order differential equation with the prescribed singularities can be written as 
\be
\label{diffeqnspecific}
\left[\partial_z^2 + \frac12\left(\frac{\Delta_1}{z^2} + \frac{\Delta_2}{(z-1)^2}+\frac{\Delta_3-\Delta_1-\Delta_2}{z(z-1)} \right)\right]\psi(z)= 0~,
\ee
where $\Delta_i = \frac12 + \frac{\lambda_i^2}{2}$. 
The local behavior of the two linearly independent solutions near each of the singularities is given by 
\be 
\left. \psi(z)\right|_{z\rightarrow z_i}  \sim z^{\frac{1\pm i\lambda_i}{2}}(1+ O(z))~ \, .
\ee 
The differential equation  \eqref{diffeqnspecific} is closely related to the hypergeometric equation and we can immediately write down three pairs of solutions in which the hyperbolic monodromy is diagonal around each given singular point:
\begin{align}
\psi_1^\pm &= \frac{e^{\pm \frac{i v_1}{2}}}{\sqrt{i \lambda_1}} z^{\frac{1\pm i \lambda_1}{2}}
(1-z)^{\frac{1 \mp i \lambda_2}{2}} 
{}_2 F_1 (\frac{1 \pm i \lambda_1 \mp i \lambda_2 \pm i \lambda_3}{2}, \frac{1 \pm i \lambda_1 \mp i \lambda_2 \mp i \lambda_3}{2} ,1 \pm i \lambda_1;z)
 \\
\psi_2^\pm &= i \frac{e^{\pm \frac{i v_2}{2}}}{\sqrt{i \lambda_2}} (1-z)^{\frac{1\pm i \lambda_2}{2}}
z^{\frac{1 \mp i \lambda_1}{2}} 
{}_2 F_1 (\frac{1 \pm i \lambda_2 \mp i \lambda_1 \pm i \lambda_3}{2}, \frac{1 \pm i \lambda_2 \mp i \lambda_1 \mp i \lambda_3}{2} ,1 \pm i \lambda_2;1-z)
\nonumber \\
\psi_3^\pm &= (iz) \frac{e^{\pm \frac{i v_3}{2}}}{\sqrt{i \lambda_3}} (\frac{1}{z})^{\frac{1\pm i \lambda_3}{2}}
(1-\frac{1}{z})^{\frac{1 \mp i \lambda_2}{2}} 
{}_2 F_1 (\frac{1 \pm i \lambda_3 \mp i \lambda_2 \pm i \lambda_1}{2}, \frac{1 \pm i \lambda_3 \mp i \lambda_2 \mp i \lambda_1}{2} ,1 \pm i \lambda_3;\frac{1}{z})\, . \nonumber
\end{align}
The requirement that the doublet $\psi_1^\pm$ has a hyperbolic $SL(2,\mathbb{R})$ monodromy around the point $z=1$ conjugate to the diagonal monodromy determined by $\lambda_2$ leads to a partial determination of the complex number $e^{i v_1}$ via hypergeometric connection formulae \cite{Hadasz:2003he, Firat:2021ukc}. The number can be chosen equal to:
\begin{align}
e^{i v_1} =&
   \frac{\Gamma(1-i\lambda_1)}{\Gamma((1+i\lambda_1)}
    \sqrt{\frac{
    \Gamma \left(\frac{1}{2}  \left(1+i\lambda _1-i\lambda _2-i\lambda _3\right)\right) 
    \Gamma \left(\frac{1}{2}  \left(1+i\lambda _1+i\lambda _2-i\lambda _3\right)\right)}
     {\Gamma \left(\frac{1}{2} \left(1-i\lambda _1+i\lambda _2+i\lambda _3\right)\right) 
    \Gamma \left(\frac{1}{2}  \left(1-i\lambda _1-i\lambda _2+i\lambda _3\right)\right)}}
    \cr 
    &\hspace{1.7cm}\times\sqrt{\frac{
    \Gamma \left(\frac{1}{2}  \left(1+i\lambda _1-i\lambda _2+i\lambda _3\right)\right) 
    \Gamma \left(\frac{1}{2}  \left(1+i\lambda _1+i\lambda _2+i\lambda _3\right)\right)}
    {\Gamma \left(\frac{1}{2} \left(1-i\lambda _1+i\lambda _2-i\lambda _3\right)\right) 
    \Gamma \left(\frac{1}{2} \left(1-i\lambda _1-i\lambda _2-i\lambda _3\right)\right)}}~.
    \label{expv1}
\end{align}
Many more details on the monodromy matrices are exhibited in Appendix \ref{AppendixHyperbolicMonodromies}. 
Let us further consider the solutions around $z=0$ and rewrite the ratio of the two independent solutions in a simplified form. By using the solution \eqref{expv1} for the phase $e^{i v_1}$, we obtain 
\begin{multline}
    \frac{\psi_1^+(z)}{\psi_1^-(z)} = z^{i \lambda_1}
     \frac{\Gamma(1-i\lambda_1)}{\Gamma((1+i\lambda_1)}
    \sqrt{\frac{
    \Gamma \left(\frac{1}{2}  \left(1+i\lambda _1-i\lambda _2-i\lambda _3\right)\right) 
    \Gamma \left(\frac{1}{2}  \left(1+i\lambda _1+i\lambda _2-i\lambda _3\right)\right)}
     {\Gamma \left(\frac{1}{2} \left(1-i\lambda _1+i\lambda _2+i\lambda _3\right)\right) 
    \Gamma \left(\frac{1}{2}  \left(1-i\lambda _1-i\lambda _2+i\lambda _3\right)\right)}}
    \cr 
    \times\sqrt{\frac{
    \Gamma \left(\frac{1}{2}  \left(1+i\lambda _1-i\lambda _2+i\lambda _3\right)\right) 
    \Gamma \left(\frac{1}{2}  \left(1+i\lambda _1+i\lambda _2+i\lambda _3\right)\right)}
    {\Gamma \left(\frac{1}{2} \left(1-i\lambda _1+i\lambda _2-i\lambda _3\right)\right) 
    \Gamma \left(\frac{1}{2} \left(1-i\lambda _1-i\lambda _2-i\lambda _3\right)\right)}}\\
\times    
\frac{{}_2 F_1\left(\frac{1}{2}  \left(1+i\lambda _1-i\lambda _2-i\lambda _3\right),
\frac{1}{2}  \left(1+i\lambda _1-i\lambda _2+i\lambda _3\right),1+i \lambda _1,z\right)}{{}_2 F_1\left(\frac{1}{2} \left(1-i\lambda _1-i\lambda _2+i\lambda _3\right),\frac{1}{2}  \left(1-i\lambda _1-i\lambda _2-i\lambda _3\right),1-i \lambda _1,z\right)} \, .
\label{AllHyperbolic}
\end{multline}
We introduce the notation 
\be
\widetilde{F}(a,b,c;z) = \frac{\Gamma(a)\Gamma(b)}{\Gamma(c)}\,  {}_2 F_1(a,b,c;z)~,
\ee
in terms of which the ratio reads 
\begin{align}
    \frac{\psi_1^+(z)}{\psi_1^-(z)} = z^{i \lambda_1}~ \tanh\frac{\gamma}{2}~ \frac{\widetilde F\left(\frac{1}{2}  \left(1+i\lambda _1-i\lambda _2-i\lambda _3\right),
\frac{1}{2}  \left(1+i\lambda _1-i\lambda _2+i\lambda _3\right),1+i \lambda _1,z\right)}{\widetilde  F\left(\frac{1}{2} \left(1-i\lambda _1-i\lambda _2+i\lambda _3\right),\frac{1}{2}  \left(1-i\lambda _1-i\lambda _2-i\lambda _3\right),1-i \lambda _1,z\right)}~,
\end{align}
where the constant factor is determined by the constraint
\be 
\left(\tanh\frac{\gamma}{2}\right)^2 = \frac{\cosh \left(\pi  \left(\lambda _1-\lambda _2\right)\right)+\cosh \left(\pi  \lambda _3\right)}{\cosh \left(\pi  \left(\lambda _1+\lambda _2\right)\right)+\cosh \left(\pi  \lambda _3\right)}~.
\ee 
The constraint can be rewritten as:
\be
\cosh\gamma = \coth \left(\pi  \lambda _1\right) \coth \left(\pi  \lambda _2\right)+\cosh \left(\pi  \lambda _3\right) \text{csch}\left(\pi  \lambda _1\right) \text{csch}\left(\pi  \lambda _2\right) \label{HHHRelation}
\ee
or
\be 
\cosh\pi\lambda_3 =\cosh\gamma \sinh\pi\lambda_1 \sinh\pi\lambda_2 -\cosh\pi \lambda_1\cosh\pi\lambda_2~.  
\ee 
These results and those in the Appendix are  minor completions of the literature \cite{Hadasz:2003he, Firat:2021ukc}. 

\subsubsection{Three Elliptic Monodromies}
The case of three elliptic monodromies was solved explicitly in \cite{Welling:1997fw}. We provide more detail and perform the calculation in close analogy to the hyperbolic case \cite{Hadasz:2003he, Firat:2021ukc}. However, it is convenient to work in a formalism in which the monodromies are once more diagonal which obliges us to consider auxiliary $SU(1,1)$ doublets $\xi^\pm$. 
We therefore start from the bases:
\begin{align}
\xi_1^\pm &= \frac{e^{\pm \frac{i v_1}{2}}}{\sqrt{\alpha_1}} z^{\frac{1\pm \alpha_1}{2}}
(1-z)^{\frac{1 \mp \alpha_2}{2}} 
{}_2 F_1 (\frac{1 \pm \alpha_1 \mp \alpha_2 \pm \alpha_3}{2}, \frac{1 \pm \alpha_1 \mp \alpha_2 \mp \alpha_3}{2} ,1 \pm \alpha_1;z)
 \\
\xi_2^\pm &= i \frac{e^{\pm \frac{i v_2}{2}}}{\sqrt{\alpha_2}} (1-z)^{\frac{1\pm \alpha_2}{2}}
z^{\frac{1 \mp \alpha_1}{2}} 
{}_2 F_1 (\frac{1 \pm \alpha_2 \mp \alpha_1 \pm \alpha_3}{2}, \frac{1 \pm \alpha_2 \mp \alpha_1 \mp \alpha_3}{2} ,1 \pm \alpha_2;1-z)
\nonumber \\
\xi_3^\pm &= (iz) \frac{e^{\pm \frac{i v_3}{2}}}{\sqrt{\alpha_3}} (\frac{1}{z})^{\frac{1\pm \alpha_3}{2}}
(1-\frac{1}{z})^{\frac{1 \mp \alpha_2}{2}} 
{}_2 F_1 (\frac{1 \pm \alpha_3 \mp \alpha_2 \pm \alpha_1}{2}, \frac{1 \pm \alpha_3 \mp \alpha_2 \mp \alpha_1}{2} ,1 \pm \alpha_3;\frac{1}{z}) \nonumber \, .
\end{align}
The  monodromies of the doublets around the corresponding singularity are diagonal $SU(1,1)$ monodromies. Demanding that the monodromies around all singular points are in the $SU(1,1)$ group, one can constrain the complex number $e^{iv_1}$ via the connection formulae and take it to be
\begin{multline}
e^{i v_1} =  \frac{\Gamma \left(1-\alpha_1\right)}{\Gamma \left(1+\alpha_1\right)}
\sqrt{
\frac{\Gamma \left(\frac{1}{2} \left(1+\alpha _1+\alpha _2+\alpha _3\right)\right)}{\Gamma \left(\frac{1}{2} \left(1-\alpha _1-\alpha _2-\alpha _3\right)\right)}
\frac{\Gamma \left(\frac{1}{2} \left(1+\alpha _1-\alpha _2+\alpha _3\right)\right)}{\Gamma \left(\frac{1}{2} \left(1-\alpha _1+\alpha _2-\alpha _3\right)\right)}} \cr
\times \sqrt{
\frac{\Gamma \left(\frac{1}{2} \left(1+\alpha _1+\alpha _2-\alpha _3\right)\right)}{\Gamma \left(\frac{1}{2} \left(1-\alpha _1-\alpha _2+\alpha _3\right)\right)}
\frac{\Gamma \left(\frac{1}{2} \left(1+\alpha _1-\alpha _2-\alpha _3\right)\right)}{\Gamma \left(\frac{1}{2} \left(1-\alpha _1+\alpha _2+\alpha _3\right)\right)}
}
\, .
\end{multline} 
The ratio of the solutions around the singularity $z=0$ equals:
\begin{align}
    \frac{\xi_1^+}{\xi_1^-} = z^{\alpha_1}\, \coth\frac{\delta}{2}\,  \frac{\widetilde{F}(\frac{1}{2} \left(1+\alpha _1-\alpha _2-\alpha _3\right),\frac{1}{2} \left(1+\alpha _1-\alpha _2+\alpha _3\right),1+\alpha _1,z)}{\widetilde{F}(\frac{1}{2} \left(1-\alpha _1-\alpha _2-\alpha _3\right),\frac{1}{2} \left(1-\alpha _1-\alpha _2+\alpha _3\right),1-\alpha _1,z)}~.
    \label{AllElliptic}
\end{align}
The prefactor  satisfies
\be 
\left(\coth\frac{\delta}{2}\right)^2 =\frac{\cos \left(\pi  \left(\alpha_1-\alpha_2\right)\right)+\cos \left(\pi  \alpha_3\right)}{\cos \left(\pi  \left(\alpha_1+\alpha_2\right)\right)+\cos \left(\pi  \alpha_3\right)}~, 
\ee
which can be rewritten as 
\be 
\cosh\delta =  \cot \left(\pi  \alpha _1\right) \cot \left(\pi  \alpha _2\right)+\cos \left(\pi  \alpha _3\right) \csc \left(\pi  \alpha _1\right) \csc \left(\pi  \alpha _2\right) \, . \label{EEERelation}
\ee
We can alternatively reparameterize elliptic monodromies $\alpha_i$ by complementary deficit angles $8 \pi G m_i$ inspired by three-dimensional gravity considerations:
\begin{align}
\alpha_{i} &= 1- 4 G m_{i} \, .
\end{align}
We can then derive the relation between the mass parameters obtained in \cite{Welling:1997fw}
\begin{align}
\cos 4 \pi G m_3 &= \cos 4 \pi G m_1 \cos 4 \pi G m_2 - \cosh \delta \sin 4 \pi G m_1 \sin 4 \pi G m_2 \, .
\end{align}
Note that these are the standard relations whose analytic continuation has been analyzed in some detail  \cite{Steif:1995pq}.
These relations are due to classical monodromy matrix multiplications in $SL(2,\mathbb{R}) \equiv SU(1,1)$. Since the monodromy matrix multiplications allow for many sub-cases depending on the nature of one or another of the three monodromies, we derive a zoo of relations between parameters, of which (\ref{EEERelation}) and (\ref{HHHRelation}) are but two examples.

\subsubsection{Two Elliptic and One Hyperbolic Monodromies}
Next, we consider the case in which we have  elliptic monodromies at the singular points $z=0$ and $z=1$  and a hyperbolic monodromy at $z=\infty$. The local solutions around each point are given by
\begin{align}
    \xi_1^\pm &= \frac{e^{\pm \frac{i v_1}{2}}}{\sqrt{\alpha_1}} z^{\frac{1\pm \alpha_1}{2}}
(1-z)^{\frac{1 \mp \alpha_2}{2}} 
{}_2 F_1 (\frac{1 \pm \alpha_1 \mp \alpha_2 \pm i \lambda_3}{2}, \frac{1 \pm \alpha_1 \mp \alpha_2 \mp i \lambda_3}{2} ,1 \pm \alpha_1;z)
 \\
\xi_2^\pm &= i \frac{e^{\pm \frac{i v_2}{2}}}{\sqrt{\alpha_2}} (1-z)^{\frac{1\pm \alpha_2}{2}}
z^{\frac{1 \mp \alpha_1}{2}} 
{}_2 F_1 (\frac{1 \pm \alpha_2 \mp \alpha_1 \pm i \lambda_3}{2}, \frac{1 \pm \alpha_2 \mp \alpha_1 \mp i \lambda_3}{2} ,1 \pm \alpha_2;1-z)
\nonumber \\
\psi_3^\pm &= (iz) \frac{e^{\pm \frac{i v_3}{2}}}{\sqrt{i \lambda_3}} (\frac{1}{z})^{\frac{1\pm i \lambda_3}{2}}
(1-\frac{1}{z})^{\frac{1 \mp \alpha_2}{2}} 
{}_2 F_1 (\frac{1 \pm i \lambda_3 \mp \alpha_2 \pm \alpha_1}{2}, \frac{1 \pm i \lambda_3 \mp \alpha_2 \mp \alpha_1}{2} ,1 \pm i \lambda_3;\frac{1}{z}) \nonumber \, .
\end{align}
Using the transformation formulas for the hypergeometric functions near $z=\infty$ to the one around $z=0$, we fix the phase number ambiguity in front of the wave-functions $\psi_3^\pm$:
\begin{multline}
e^{i v_3} =e^{2\pi i \alpha_2}
\frac{\Gamma(1-i \lambda_3)}{\Gamma(1+i \lambda_3)}
\sqrt{
\frac{\Gamma \left(\frac{1}{2} \left(1-\alpha _1-\alpha _2+i \lambda _3\right)\right)}{\Gamma \left(\frac{1}{2} \left(1-\alpha _1-\alpha _2-i \lambda _3\right)\right)}
\frac{\Gamma \left(\frac{1}{2} \left(1+\alpha _1-\alpha _2+i \lambda _3\right)\right)}{\Gamma \left(\frac{1}{2} \left(1+\alpha _1-\alpha _2-i \lambda _3\right)\right)}
}
\\
\times \sqrt{
\frac{\Gamma \left(\frac{1}{2} \left(1-\alpha _1+\alpha _2+i \lambda _3\right)\right)}{\Gamma \left(\frac{1}{2} \left(1-\alpha _1+\alpha _2-i \lambda _3\right)\right)}
\frac{\Gamma \left(\frac{1}{2} \left(1+\alpha _1+\alpha _2+i \lambda _3\right)\right)}{\Gamma \left(\frac{1}{2} \left(1+\alpha _1+\alpha _2-i \lambda _3\right)\right)}
}
\end{multline}
The ratio of the solutions around $z=\infty$ is  
\begin{equation}
    \frac{\psi_3^{+}}{\psi_3^-}= e^{2\pi i \alpha_2}\, 
    {z^{-i\lambda_3}}\, 
    e^{i \epsilon}\, \frac{\widetilde{F}(\frac{1}{2}(1 + i \lambda_3 - \alpha_2 + \alpha_1), \frac{1}{2}(1 + i \lambda_3 - \alpha_2 - \alpha_1) ,1 + i \lambda_3;\frac{1}{z}) }{\widetilde{F}(\frac{1}{2}(1 - i \lambda_3 - \alpha_2 +\alpha_1), \frac{1}{2}(1 - i \lambda_3 - \alpha_2 - \alpha_1) ,1 - i \lambda_3;\frac{1}{z}) }~, \label{EEHSolution}
\end{equation}
where the square of the phase factor satisfies
\be
\label{EEHrelation}
e^{2i\epsilon} = \frac{\cos \left(\pi  \alpha _1\right)+\cos \left(\pi  \left(\alpha _2-i \lambda _3\right)\right)}{\cos \left(\pi  \alpha _1\right)+\cos \left(\pi  \left(\alpha _2+i \lambda _3\right)\right)}~.
\ee
An interesting rewriting of the relation is 
\be
\cos\pi\alpha_1 = \cot (\epsilon )\sin \left(\pi  \alpha _2\right) \sinh \left(\pi  \lambda _3\right) -\cos \left(\pi  \alpha _2\right) \cosh \left(\pi  \lambda _3\right) \, .
\ee

\subsubsection{One Elliptic and Two Hyperbolic Monodromies}
For the case of one elliptic and two hyperbolic monodromies, we set the elliptic singularity at the  point $z=1$. The local doublets around the three points are:
\begin{align}
\psi_1^\pm &= \frac{e^{\pm \frac{i v_1}{2}}}{\sqrt{i \lambda_1}} z^{\frac{1\pm i \lambda_1}{2}}
(1-z)^{\frac{1 \mp \alpha_2}{2}} 
{}_2 F_1 (\frac{1 \pm i \lambda_1 \mp \alpha_2 \pm i\lambda_3}{2}, \frac{1 \pm i \lambda_1 \mp \alpha_2 \mp i\lambda_3}{2} ,1 \pm i \lambda_1;z)
 \\
\xi_2^\pm &= i \frac{e^{\pm \frac{i v_2}{2}}}{\sqrt{\alpha_2}} (1-z)^{\frac{1\pm \alpha_2}{2}}
z^{\frac{1 \mp i \lambda_1}{2}} 
{}_2 F_1 (\frac{1 \pm \alpha_2 \mp i \lambda_1 \pm i\lambda_3}{2}, \frac{1 \pm \alpha_2 \mp i \lambda_1 \mp i\lambda_3}{2} ,1 \pm \alpha_2;1-z)
\nonumber \\
\psi_3^\pm &= (iz) \frac{e^{\pm \frac{i v_3}{2}}}{\sqrt{i\lambda_3}} (\frac{1}{z})^{\frac{1\pm i\lambda_3}{2}}
(1-\frac{1}{z})^{\frac{1 \mp \alpha_2}{2}} 
{}_2 F_1 (\frac{1 \pm i\lambda_3 \mp \alpha_2 \pm i \lambda_1}{2}, \frac{1 \pm i\lambda_3 \mp \alpha_2 \mp i \lambda_1}{2} ,1 \pm i\lambda_3;\frac{1}{z}) \, . \nonumber
\end{align}
Compatibility of the monodromies narrows down the complex pre-factors. For instance, we can pick
\begin{multline}
    e^{i v_1} =  \frac{\Gamma \left(1-i \lambda _1\right)}{\Gamma \left(1+i \lambda _1\right)}
    \sqrt{
    \frac{\Gamma \left(\frac{1}{2} \left(1-\alpha _2+i \lambda _1-i \lambda _3\right)\right)}{\Gamma \left(\frac{1}{2} \left(1-\alpha _2-i \lambda _1+i \lambda _3\right)\right)}
    \frac{\Gamma \left(\frac{1}{2} \left(1+\alpha _2+i \lambda _1-i \lambda _3\right)\right)}{\Gamma \left(\frac{1}{2} \left(1+\alpha _2-i \lambda _1+i \lambda _3\right)\right)}
    }\cr
    \times \sqrt{
    \frac{\Gamma \left(\frac{1}{2} \left(1-\alpha _2+i \lambda _1+i \lambda _3\right)\right)}{\Gamma \left(\frac{1}{2} \left(1-\alpha _2-i \lambda _1-i \lambda _3\right)\right)}
    \frac{\Gamma \left(\frac{1}{2} \left(1+\alpha _2+i \lambda _1+i \lambda _3\right)\right)}{\Gamma \left(\frac{1}{2} \left(1+\alpha _2-i \lambda _1-i \lambda _3\right)\right)}
    }
    \, .
\end{multline}
The ratio of the solutions around $z=0$ equals
\begin{align}
    \frac{\psi_1^+}{\psi_1^-} = z^{i\lambda_1}\, e^{i\eta}\, \frac{\widetilde{F}\left(\frac{1}{2} (1-\alpha_2+ i\lambda_1-i\lambda_3),\frac{1}{2} (1-\alpha_2+ i\lambda_1+i\lambda_3),1+i \lambda _1,z\right)}{\widetilde{F}\left(\frac{1}{2} (1-\alpha_2 - i\lambda_1+i\lambda_3),\frac{1}{2} (1-\alpha_2- i\lambda_1-i\lambda_3),1-i \lambda _1,z\right)}~, \label{HEHSolution}
\end{align}
where the square of the phase factor obeys
\be 
\label{etadefn}
e^{2i\eta} = \frac{\cosh \left(\pi  \lambda _3\right)+\cosh \left(\pi  \left(\lambda _1+i \alpha _2\right)\right)}{\cosh \left(\pi  \lambda _3\right)+\cosh \left(\pi  \left(\lambda _1-i \alpha _2\right)\right)} \, .
\ee 
One can rewrite this equation using the complementary parameter $\alpha_2 =1-4 G m_2 $ to express the hyperbolic monodromy $\lambda_3$ as 
\be 
\cosh\pi\lambda_3 =\cot (\eta ) \sinh \left(\pi  \lambda _1\right)  \sin \left(4 \pi  G m_2\right)+\cosh \left(\pi  \lambda _1\right) \cos \left(4 \pi  G m_2\right) \, .
\ee 
Thus, we have constructed the four types of  solutions in technical detail. Each case has its own specificity.

\subsection{The Outermost Geodesics }
\label{GeodesicsAndSingularities}

\begin{figure}[H]
\begin{center}
\includegraphics[width=8cm]{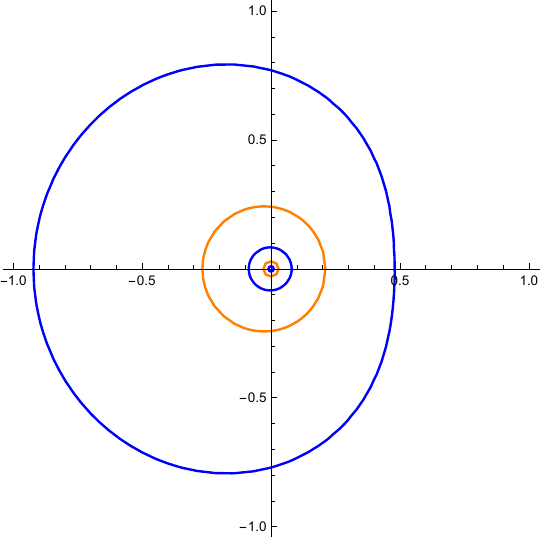}
\end{center}
\caption{  Alternating geodesics and contours around the singularity at $z=0$ in the case of the triple hyperbolic solution with $\lambda_i=1.5$. The outermost contour is the farthest geodesic before we encounter another singular point. Alternating inward we have singularities in orange and further geodesics in blue.  }
\label{Alternating}
\end{figure}

Around each hyperbolic singularity, we can draw a contour diagram which consists of closed loops around the singularity at fixed $|\rho|$. From the form of the metric (\ref{MetricNearHyperbolic}) we surmise that these closed loops are geodesics when \cite{Hadasz:2003kp,Hadasz:2003he,Firat:2021ukc}
\begin{equation}
|\rho| = e^{\frac{\pi}{\lambda} (l_g+\frac{1}{2})}
\end{equation}
with $l_g \in \mathbb{Z}$. In addition there are  boundaries at which the metric diverges when 
\begin{equation}
|\rho| = e^{\frac{\pi}{\lambda} l_b}
\end{equation}
with $l_b \in \mathbb{Z}$. These two types of contours alternate around a given singularity \cite{Hadasz:2003kp,Hadasz:2003he,Firat:2021ukc}. See the contour plot in Figure \ref{Alternating} for an example.
\noindent
The alternating pattern repeats indefinitely as we approach the singularity, but it ends as we go further away. The process ends because we do not allow for contours that circle not only the hyperbolic singularity under study but another singularity as well i.e. we fix the homotopy class of the geodesic to be equivalent to the loop around the hyperbolic singularity only.   When the influence of the other singularities becomes  important, the pattern ends and we find an outermost geodesic that surrounds a given hyperbolic singularity \cite{Hadasz:2003he, Firat:2021ukc}. The outermost geodesic is labeled by $l_g^0$ and the singular boundary that lies just inside is labeled by $l_b^0$.  

\begin{figure}[H]
\begin{center}
\includegraphics[width=10cm]{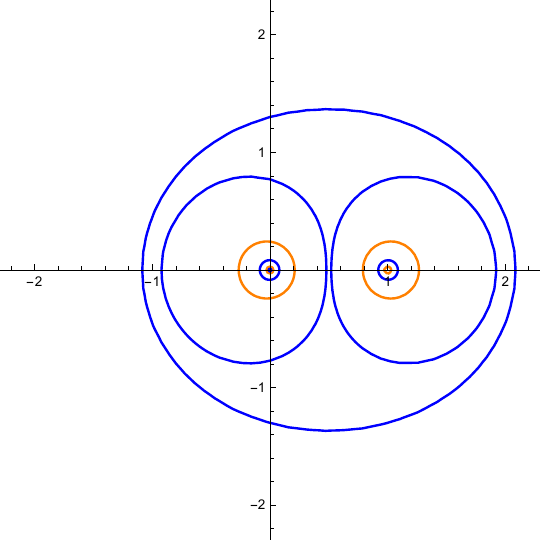}
\end{center}
\caption{  Alternating geodesics and contours around the singularities at $z=0, 1, \infty$  in the case of the triple hyperbolic solution with $\lambda_i=1.5$. The outermost contour is the farthest geodesic from $z=\infty$. Outside this curve, there will be an infinite sequence of alternating singular curves and geodesics  around infinity.}
\label{allgeoandsing}
\end{figure}

\begin{figure}[ht]
\begin{center}
\includegraphics[width=12cm]{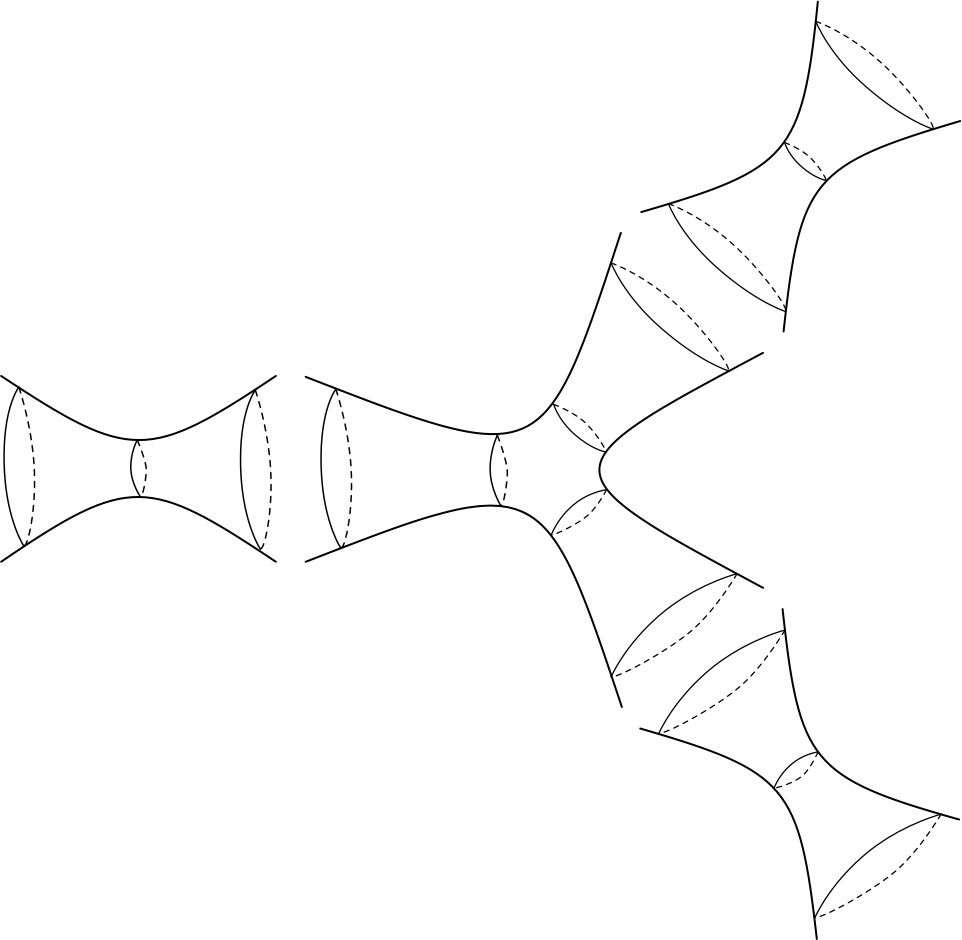}
\end{center}
\caption{A sketch of the two-dimensional geometry with three hyperbolic monodromies. Around each singularity, there is an infinite sequence of hyperbolas. They are joined by a pants diagram.}
\label{PantsWithDiabolos}
\end{figure}

For three hyperbolic singularities, the resulting geometry was described in detail in \cite{Hadasz:2003he, Firat:2021ukc} and  we exhibit it in Figure \ref{allgeoandsing}. 
\noindent
We have a middle  region bordered by the farthest geodesic around each of the hyperbolic singularities. It takes the form of a pants diagram. We can attach hyperboloids to the legs such that they end in a widening boundary which blows up at the location of the singularity $l_b^0$. That represents the maximal regular extension of the two-dimensional geometry. 
One way in which to understand the geometry is as follows. Suppose we have a closed loop in the homotopy class corresponding to circling a single hyperbolic singularity. We want to determine a geodesic in this class. The idea is that away from a singular curve, the loop shrinks. 
We have a minimization problem for a given homotopy class. The curve cannot slip off to a singular curve because there the length of the curve blows up -- it is the opposite direction one wants to take when minimizing.  For the problem with three hyperbolic singularities, it is clear that the solution will not slip off all the way to another hyperbolic singularity since we will encounter once more horns of trumpets. Thus, there is a solution to the minimization problem. This is confirmed numerically.\footnote{A proof is suggested in \cite{Firat:2021ukc}.} The resulting geometry is sketched in Figure \ref{PantsWithDiabolos}.  

\begin{figure}[H]
\begin{center}
\includegraphics[width=8cm]{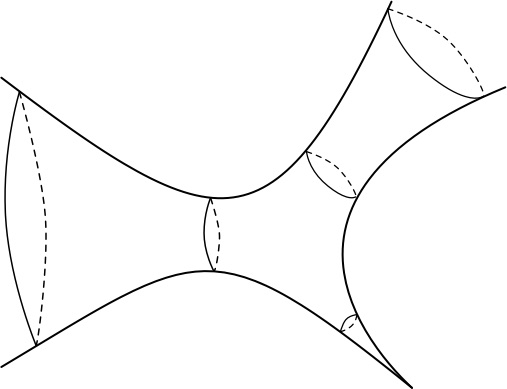}
\end{center}
\caption{A sketch of the geometry with one elliptic and two hyperbolic singularities. The pants diagram has one leg that is closed up by a cone. Infinite sequences of hyperbolas surrounding the hyperbolic singularities are not drawn.}
\label{PantsWithCone}
\end{figure}

\begin{figure}[H]
\begin{center}
\includegraphics[width=8cm]{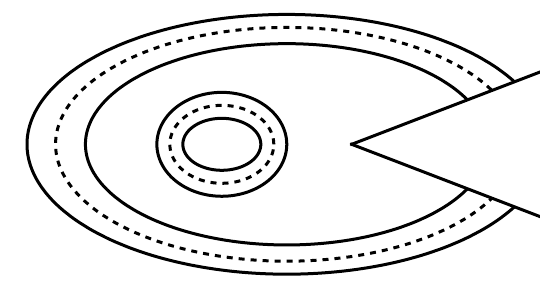}
\end{center}
\caption{We schematically represented a geometry with an elliptic singularity that cuts out a wedge on the right as well as a hyperbolic singularity surrounded by alternating geodesics and singularities. At infinity as well, there is a hyperbolic singularity. The straight lines on the right are identified.}
\label{Fish}
\end{figure}

When we replace one hyperbolic source by an elliptic source, we expect the picture to change as follows. Around the elliptic source the two-dimensional geometry is akin to a cone. Around a hyperbolic singularity, the alternating geodesic-boundary picture remains valid. 
If there is an elliptic singularity, we can potentially drag the closed curve all the way until it touches an elliptic singularity. 
Still, this behavior is not expected for a geodesic since it is favorable, close to an elliptic, conical space, to make the curve shorter by not visiting the region near the singularity at all. Thus, we expect the property that there is an outermost geodesic in the homotopy class associated to a loop circling a hyperbolic singularity  to be valid for geodesics around any hyperbolic singularity independently of the nature of the other singularities.   Therefore, if we have for instance two hyperbolic and one elliptic singularity, we expect the geometry to take the form of Figure \ref{PantsWithCone}
if we embed it in three-dimensional space. 
\noindent
One leg of the (all-hyperbolic) pants diagram has been replaced by the cone geometry that surrounds an elliptic singularity. In terms of the alternating pattern of singularities around hyperbolic geometries and the wedge insertion, we expect a two-dimensional geometry as schematically depicted in Figure \ref{Fish}. 
\noindent
%
%

\begin{figure}[H]
\begin{center}
\includegraphics[width=8cm]{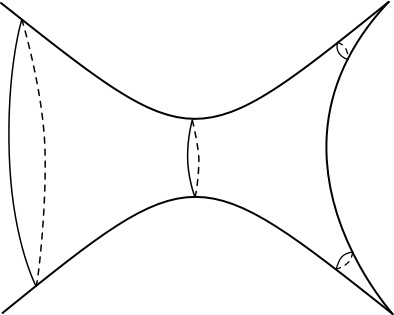}
\end{center}
\caption{
A sketch of the geometry with one hyperbolic singularity on the left and two elliptic singularities on the right. The former has an associated trumpet horn geometry (with hyperbolas attached) and the latter close up in a cone with a tip at the singularity.}
\label{2horns}
\end{figure}

\begin{figure}[H]
\begin{center}
\includegraphics
[width=6cm]
{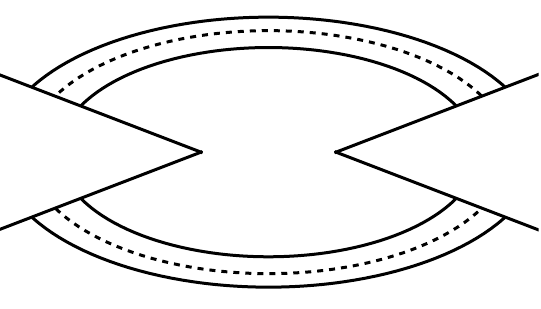}
\end{center}
\caption{There are two elliptic singularities that cut out a wedge from space. The lines bordering a wedge are identified.   There is a hyperbolic singularity at infinity which is surrounded by geodesics and singular boundaries.}
\label{EEHLogo}
\end{figure}

Similar arguments suggest the schematic pictures in 
 Figures \ref{2horns} and \ref{EEHLogo} for the case of one hyperbolic and two elliptic singularities. 
\noindent
The figures schematically represent the geometry embedded in a three-dimensional Euclidean space as well as a sketch of the wedge cut out near an elliptic singularity and the alternating pattern of geodesics and blow-ups near hyperbolic singularities.  
\noindent
 Numerically, we find that the outermost geodesic is always at $l_g^0=-1$ (with the choice of $\rho_i$ given previously). This was noted in \cite{Firat:2021ukc} for the triple hyperbolic case and seemingly continues to hold for the mixed set-ups. 
%
We stress that the observations in this subsection are preliminary. It is desirable  to find a proof of our proposed pictures. 

\section{Large Hyperbolic Monodromies and Their Saddles}
\label{Saddles}
\label{LimitSolutions}
\label{LargeHyperbolicMonodromy}
In this section, we study limits of the classical Liouville solutions. We restrict ourselves to studying limits in which the hyperbolic parameters $\lambda_i$ tend to infinity in a specified manner. 
These limits are relevant to string field theory \cite{Costello:2019fuh,Firat:2021ukc} as well as three-dimensional gravity (as we show in section \ref{3dGravity}). The limits have the advantage of leading to simpler expressions for the functions $\rho_i$ and therefore for the two-dimensional metric. They also allow to get a better handle on the geodesics.

The general technique for evaluating the limit of the Liouville field is to represent the hypergeometric functions that occur in the solutions in terms of integrals and to approximate the integrals uniformly using a saddle point approximation. The latter approximations were originally studied by Riemann \cite{Riemann} and since then ever more systematically, including in  recent and relevant contributions \cite{ParisII,CFU,Watson,Jones,KhawajaDaalhuis}.
The integrals we study will be of the form
\begin{equation}
I(z) = \int_{a}^b dt~ \chi(t,z)~ e^{\kappa \sigma(t,z)}~
\end{equation}
with $\kappa$ a large parameter. For generic values of the parameter $z$, the integral may localize on a dominant saddle point. Near special values of  $z$, contributions can localize on  coalescing saddle points or other special features of the integrand like branch points or the ends of the integration region. When the asymptotic expansion is valid for a region of values of $z$, it is called uniform and our expansions will be of that type.  Many different types of behavior have been identified and feature in limits of the classical Liouville solutions. The zoology of possibilities is too extended to review, but we shall indicate relevant features of the examples we encounter. 

\subsection{Three Large and Equal Hyperbolic Monodromies}
If we concentrate on the all-hyperbolic solution, then an interesting special case is where  the hyperbolic parameters $\lambda_i$ are all equal to one large parameter $\lambda$. The limit solution (at leading order in $1/\lambda$) was discussed in  \cite{Firat:2021ukc} where a Taylor expansion (in the position coordinate $z$) of the solution was matched to the analytic expression for the hyperbolic three string vertex \cite{Costello:2019fuh,Firat:2021ukc}. 
Here, we will derive the string vertex by a uniform approximation to the hypergeometric function using a saddle point technique. Moreover, we show that we control $1/\lambda^2$ corrections to the metric and provide a detailed description of what happens near the double saddle point. This allows us to derive an accurate picture of the behavior of the farthest geodesics.

\subsubsection{The Approximation to the Metric}
We consider the function $\rho_1$ \eqref{fdefnz0} that is single-valued around the $z=0$ singularity: 
\be
\rho_1 = \left(\frac{\psi^+_1(z)}{\psi^-_1(z)}\right)^{\frac{1}{i\lambda_1}}~,
\ee 
where the doublet is the all-hyperbolic solution determined in subsection \ref{AllHyperbolicSubsection}
and we consider the situation in which we set all the $\lambda_i$ to be equal to $\lambda$, and then take the large $\lambda$ limit. Explicitly, for the case of equal $\lambda_i$, we have
\begin{align}
    \rho_1 
%
&= e^{\frac{v_1}{\lambda}}\, z\, (1-z)\,   \left(\frac{{}_2 F_1 (\frac{1 + i \lambda}{2}, \frac{1  +3 i \lambda}{2} , 1 +i \lambda;z)}{{}_2 F_1 (\frac{1  - i \lambda}{2}, \frac{1  -3 i \lambda}{2} ,1 - i \lambda;z)}\right)^{\frac{1}{i\lambda}} \, .
\end{align}
We have used the Euler identity for the hypergeometric functions. 

At this stage,  we pause to review the asymptotic series expansion for the hypergeometric function in the limit of large parameter. Following the notations of \cite{ParisII} we define 
\be 
F_3(\kappa;z) := {}_2 F_1 (a+\epsilon_1\kappa, b + \epsilon_2\kappa, c+\kappa; z )~. 
\ee
Our formulas contain functions of this type with parameters:
\begin{equation}
a=\frac{1}{2}=b \, , \qquad c=1 \, , \qquad \epsilon_1=\frac{1}{2} \, , \qquad  \epsilon_2=\frac{3}{2} \, \qquad \kappa = \pm i \lambda \, .
\end{equation}
The large $\kappa$ expansion of the function $F_3$ is obtained via an integral representation
\be 
F_3(\kappa;z) = \frac{\Gamma(c+\kappa)}{\Gamma(a+\epsilon_1\kappa)\Gamma(c-a+(1-\epsilon_1)\kappa)}\int_0^1 dt\, \chi(t)\, e^{\kappa\, \sigma(t)}~, \label{F3Integral}
\ee 
and a saddle point analysis at large $|\kappa|$. Here the functions $\sigma$ and $\chi$ take the form
\begin{align}
    \sigma(t) &= \epsilon_1\log t+(1-\epsilon_1)\log(1-t)-\epsilon_2\log(1-zt)~, \quad \chi(t) = \frac{t^{a-1}(1-t)^{c-a-1}}{(1-zt)^b}~.
\end{align}
We refer the reader to \cite{ParisII} for many more details and  present only the  results. The saddle points $t_s$ of the integral lie at
\be 
t_{sj} = \frac{2\epsilon_1}{\Upsilon_{\pm}}~, \quad\text{for}\quad j=1,2~,
\ee 
where the upper (lower) sign is for $j=1$ ($j=2$) and we have introduced the quantities
\begin{align}
    \Upsilon_{\pm} = \Delta \pm \sqrt{\Delta^2-4\epsilon_1(1-\epsilon_2)z}~ \quad\text{with}\quad \Delta = 1+(\epsilon_1-\epsilon_2)z~.
\end{align}
The asymptotic series  in the regime 
\begin{equation}
\text{Re} \, \kappa (\sigma(t_{s1})-\sigma(t_{s2})) > 0
\end{equation}
is obtained from the saddle point $t_{s1}$. We suppose we are in this regime from now on.
The series is given by
\begin{align}
    F_3(\kappa;z) = \frac{2\pi i \Gamma(c+\kappa)}{\Gamma(a+\epsilon_1\kappa)\Gamma(c-a+(1-\epsilon_1)\kappa)}\frac{\chi(t_{s1}) e^{\kappa \sigma(t_{s1})}}{\sqrt{2\pi \sigma''(t_{s1})}}\sum_{n=0}^{\infty} \frac{c_{n}^{(1)} \Gamma(n+\frac12)}{\kappa^{n+\frac12}\Gamma(\frac12)}~.
\end{align}
The coefficients $c_{n}^{(1)}$ in this series are given in terms of the derivatives of the exponent $\sigma(t)$ evaluated at the saddle point $t_{s1}$. The first three are:
\begin{align}
    c_0^{(1)}=1~, \quad c_1^{(1)} = -\frac{1}{\sigma''}\left(\frac{\chi''}{\chi}-\frac{\sigma'''}{\sigma''} \frac{\chi'}{\chi}+\frac14\left(\frac53\frac{(\sigma''')^2}{(\sigma'')^2}-\frac{\sigma^{(4)}}{\sigma''}\right) \right)~,
\end{align}
and 
\begin{align}
c_2^{(1)} =&\frac{1}{(\sigma '')^2}\Bigg\{
\left(\frac{\chi^{(4)}}{6 \chi}-\frac{5 \chi^{(3)} \sigma ^{(3)}}{9 \chi \sigma''}+\frac{5}{12} \left(\frac{7}{3} \left(\frac{\sigma ^{(3)}}{\sigma ''}\right)^2-\frac{\sigma ^{\text{(4)}}}{\sigma ''}\right) 
\frac{\chi''}{\chi} \right)\cr
&-\frac{35}{36}\frac{\chi'}{\chi}\left(-\frac{\sigma ^{(3)} \sigma ^{\text{(4)}}}{\left(\sigma ''\right)^2}+\frac{6 \sigma^{(5)}}{35 \sigma ''}+\left(\frac{\sigma ^{(3)}}{\sigma ''}\right)^3\right)\cr
&+\frac{35}{36}\left[\frac{11}{24} \left(\frac{\sigma ^{(3)}}{\sigma ''}\right)^4+\frac{1}{5}\frac{ \sigma ^{(3)} \sigma ^{(5)}}{\left(\sigma ''\right)^2}-\frac{\sigma ^{\text{(6)}}}{35 \sigma ''} -\frac{3}{4}\frac{\sigma ^{\text{(4)}}}{\sigma ''} 
\left(\left(\frac{\sigma ^{(3)}}{\sigma ''}\right)^2-\frac{\sigma ^{\text{(4)}}}{6 \sigma ''}\right)
\right]
\Bigg\}~.
\end{align}
Using these results for the asymptotic expansion of the hypergeometric function and the result in equation \eqref{expv1} for the complex phase $e^{i v_1}$, the leading term in the large $\lambda$ expansion is 
\be 
\label{rho1defn}
\rho_1 = e^{- \frac{\pi}{2 \lambda}} \, e^{\frac{v_1}{\lambda}}\,  \left( \rho_1^{(0)}(z) + \frac{1}{\lambda^2}\rho_1^{(2)}(z) + \ldots\right) ~,
\ee 
where the pre-factor is 
\begin{align}
\label{expv1bylambda}
    e^{\frac{v_1}{\lambda}} &=
    \left(\frac{\Gamma \left(\frac{1}{2}-\frac{i \lambda }{2}\right)}{\Gamma \left(\frac{1}{2}+\frac{i \lambda }{2}\right)} \right)^{-\frac{3 i}{2 \lambda }} 
\left(\frac{\Gamma \left(\frac{1}{2}+\frac{3 i \lambda }{2}\right)}{\Gamma \left(\frac{1}{2}-\frac{3 i \lambda }{2}\right)}\right)^{-\frac{i}{2 \lambda }}
    \nonumber \\
    &\approx 3\sqrt{3} e^{- \frac{2}{9 \lambda^2}}  
    \approx 3\sqrt{3} -\frac{2}{\sqrt{3} \lambda ^2} + O(\lambda^{-4})~,
\end{align} 
and the first term is given by\footnote{The leading term confirms an educated guess in \cite{Firat:2021ukc}.}
\begin{align}
\label{rho10}
    \rho_1^{(0)} &= \frac{(1-z) \left(1-\sqrt{(z-1) z+1}\right)}{\left(-z-\sqrt{(z-1) z+1}+2\right)^3 \left(-z+\sqrt{(z-1) z+1}+1\right)}~.
\end{align}
The sub-leading term is cumbersome, but expanding the product in equation \eqref{rho1defn} up to order $\lambda^{-2}$, we find 
\begin{align}
    \rho_1 =  3\sqrt{3}\, e^{-\frac{\pi}{2\lambda}} \, \rho_1^{(0)}\left(1+\frac{5 (z-2) (z+1) (2 z-1)}{36  ((z-1) z+1)^{\frac{3}{2}}}\frac{1}{\lambda ^2} + \ldots \right)
    \label{rho1atorderlambdasquared}
\end{align}
The maximal validity of the expression is limited by the branch points of the square root. They are on the unit circle at angles $\pm \pi/3$ with respect to the positive real axis in the $z$-plane.
These approximations are largely uniform in the argument $z$. The second order correction has interesting properties. Near the branch points, the correction term is relatively stronger. In fact, the branch points give rise to double saddle points in the saddle point integration and in their neighborhood, a different saddle point approximation method must be used \cite{CFU}. 
 We will return to this regime near the branch points shortly. 
Note  that the zeroes of the numerator of the second order correction lie at $z=-1,1/2$ and $2$. At those points, the leading order result remains valid (up to order $\lambda^{-2}$ and a universal rescaling). 

When all hyperbolic monodromies $\lambda_i$ are equal, the functions $\rho_{2,3}$ are given by $\rho_2(z)=\rho_1(1-z)$ and $\rho_3(z)=\rho_1(1/z)$ \cite{Firat:2021ukc}. The branch points of these functions are once more at $z=e^{\pm i \pi/3}$ since the pair of branch points is mapped to the same pair under $z \rightarrow 1-z$ and $z  \rightarrow 1/z$. From the definition of the functions $\rho_i$ it is clear that they will have the same expansions around their respective center $z_i$.

\subsubsection{Near  Branch Points }
Recall that we have the integral expression (\ref{F3Integral}) for the function $F_3$. The integrand has an exponent 
\be 
\sigma(t)= \epsilon_1\log t + (1-\epsilon_1)\log(1-t) - \epsilon_2\log(1-zt)~.
\ee 
The saddle points are the solutions to the equation $\sigma'(t)=0$:
\be 
t_\pm = \frac{1}{z}(z-1 \pm \sqrt{1-z+z^2})~.
\ee 
Thus, the saddle points coalesce in the $t$-plane  when 
\be 
1 - z + z^2 = 0~ \quad\text{i.e.}\quad z_b =  e^{\pm \frac{\pi i }{3}}~.
\ee 
Near e.g. the branch point $z_b=e^{\frac{\pi i}{3}}$ of the argument of the square root, our previous Gaussian expression for the asymptotic expansion breaks down and one has to perform an extended saddle point analysis \cite{CFU}.
Near the double saddle, the exponent can be transformed in a regular manner in a cubic expression. The cubic expression gives rise to an overall exponent $A$ which is related to the sum of saddle point exponents:
\begin{align}
A &=
\frac{\sigma(t_{+})+\sigma(t_{-})}{2}
=
\frac{1}{4} \log \left(-\frac{1}{27 (z-1)^2 z^2}\right)
\end{align}
and an Airy function which depends on an argument $\kappa^{\frac{2}{3}} \zeta$ defined in terms of the difference of exponents:
\begin{align}
\frac{2}{3} \zeta^{\frac{3}{2}} &=
\frac{\sigma(t_{-})-\sigma(t_{+})}{2}  \\
&=
\frac{1}{4} \log\left[ \displaystyle{\frac{\left(1-\sqrt{(z-1) z+1}\right) \left(-z+\sqrt{(z-1) z+1}+2\right)^3 \left(\sqrt{(z-1) z+1}-1+z\right)}{\left(1+\sqrt{(z-1) z+1}\right) \left(z+\sqrt{(z-1) z+1}-2\right)^3\left(\sqrt{(z-1) z+1}+1-z\right)  }}\right] \, . \nonumber
\end{align}
When $\kappa^{\frac{2}{3}} \zeta$ is of order one or smaller and to leading order in a $\kappa^{-\frac{1}{3}}$  expansion, the hypergeometric is approximated by an Airy function \cite{CFU}:
\begin{align}
F_3(\kappa, z) & \approx 2\pi i 
\frac{\Gamma(c+\kappa)}{ \Gamma(a+\epsilon_1 \kappa) \Gamma(c-a+(1-\epsilon_1) \kappa)}
e^{\kappa A}  \left(
p_0 \frac{Ai(\kappa^\frac{2}{3} \zeta)}{\kappa^\frac{1}{3}}   + O(\kappa^{-\frac{2}{3}})
 \right) 
 \, .
\end{align}
The coefficient function $\chi(t)$ determines the expansion of the integrand in the neighborhood of the saddle point contributions and therefore the coefficient function $p_0$. 

We can plug this approximation to the hypergeometric function into $\rho_1$ to find:
\begin{align}
\rho_1 = \frac{i}{3\sqrt{3}}\, 
e^{-\frac{\pi}{3\lambda}}
\left(\frac{\Gamma \left(\frac{1}{2}-\frac{i \lambda }{2}\right)}{\Gamma \left(\frac{1}{2}+\frac{i \lambda }{2}\right)} \right)^{-\frac{3 i}{2 \lambda }} 
\left(\frac{\Gamma \left(\frac{1}{2}+\frac{3 i \lambda }{2}\right)}{\Gamma \left(\frac{1}{2}-\frac{3 i \lambda }{2}\right)}\right)^{-\frac{i}{2 \lambda }} 
\left( 
\frac{ 
\text{Ai}\left(e^{\frac{\pi i}{3}}\zeta  \lambda ^{\frac{2}{3}}\right)
}
{\text{Ai}\left( e^{-\frac{\pi i}{3}}\zeta  \lambda ^{\frac{2}{3}}\right)}
\right)^{-\frac{i}{\lambda }} ~.
\label{AiryApproximation}
\end{align}
A  check of this expression is to match the leading order expression for large $\lambda$. The $\Gamma$ functions can be expanded as before in equation \eqref{expv1bylambda}, and we use the  asymptotic expansion for the Airy function 
\be 
\text{Ai}(x) \sim \frac{e^{-\frac{2}{3}x^{\frac32} }}{2\sqrt{\pi}x^{\frac{1}{4}}}\left(1-\frac{5}{48x^{\frac32}} + \ldots \right)~. 
\ee 
We then obtain 
\be 
\rho_1 = i\, e^{-\frac{\pi}{2\lambda}}\, e^{-\frac43 \zeta^{\frac32}} \left(1 + O(\lambda^{-2} ) \right)~.
\ee 
Substituting for $\zeta$, we recover the leading order result $\rho_1^{(0)}$ in equation \eqref{rho10}. 
It also interesting to consider a large $\lambda$ limit at fixed
 $\zeta \lambda^\frac{2}{3}$. This is very close to the branch point but in a regime where the Airy function argument is of order one and the distance to the branch point scales as $\lambda^{-\frac{2}{3}}$. We will use equation \eqref{AiryApproximation} to plot large $\lambda$ geodesics near the branch point shortly. 

\subsubsection{The  Farthest Geodesics}
\begin{figure}[H]
\begin{center}
\includegraphics[width=5.3cm]{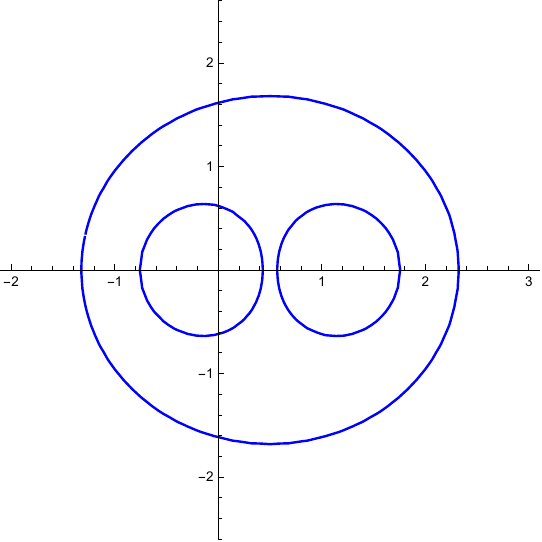}
\includegraphics[width=5.3cm]{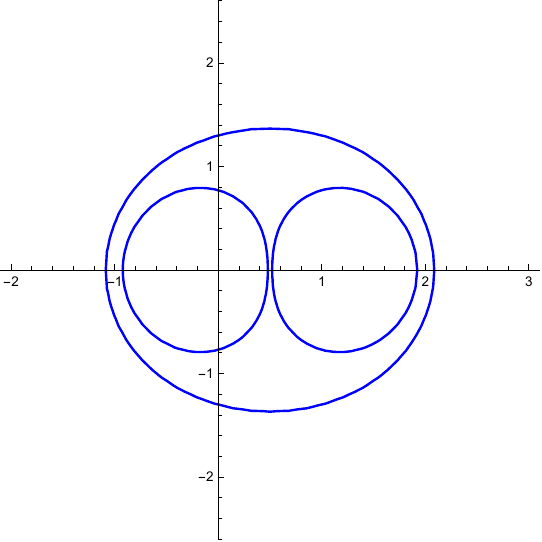}
\includegraphics[width=5.3cm]{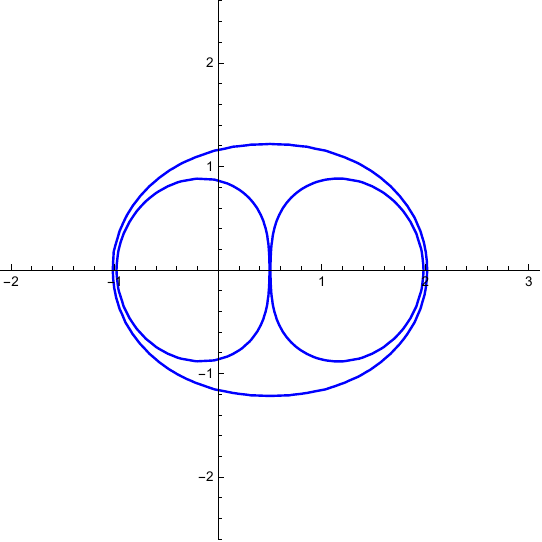}
\end{center}
\caption{The numerical outermost geodesics in the $z$-plane at various values of a common $\lambda=\lambda_i$. The values are $1, 1.5$ and $2.2$.
}
\label{HHHShortestGeodesics}
\end{figure}

\begin{figure}
\begin{center}
\includegraphics[width=5.3cm]{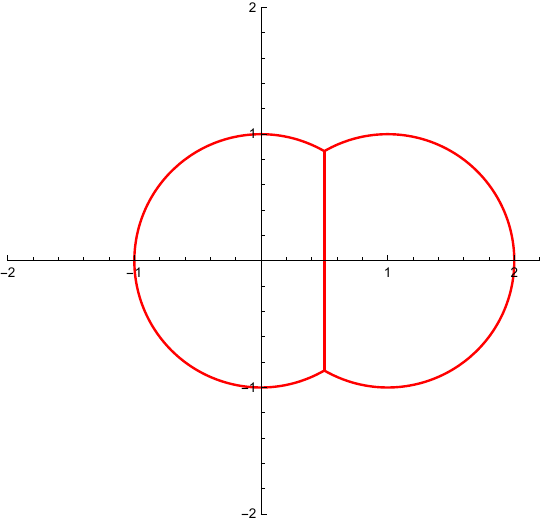}
\includegraphics[width=5.3cm]{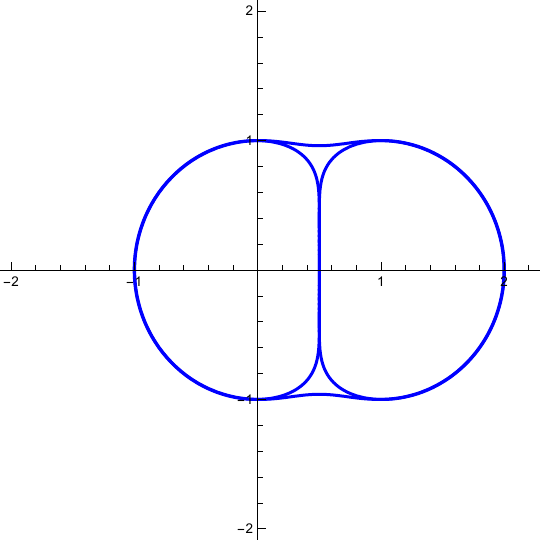}
\includegraphics[width=5.3cm]{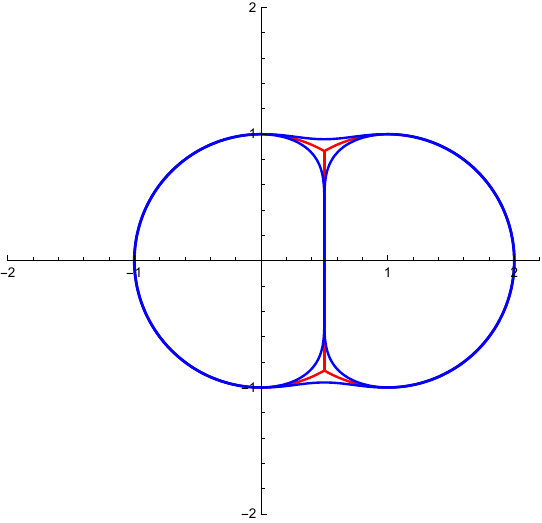}
\end{center}
\caption{The limit curves are shown on the left. The numerical outermost geodesics at a large value of a common $\lambda=12$ is shown in the middle. 
On the right, the overlay of the two figures. The parameter $\lambda$ is sufficiently large to find agreement away from the branch points.   
}
\label{LimitCurves}
\end{figure}

The Liouville field codes a two-dimensional metric. The three functions $\rho_{i=1,2,3}$ describe the metric in three patches near $z=0,1$ and $\infty$. As discussed in section \ref{GeodesicsAndSingularities}, we can define outermost geodesics in a given homotopy class of loops around a single singularity.  A numerical description of these curves for intermediate values of $\lambda$ is provided in Figure \ref{HHHShortestGeodesics}. See also \cite{Hadasz:2003he} as well as \cite{Firat:2021ukc} for very similar graphs. At large values of $\lambda$ the numerics becomes challenging. 
We can also obtain a precise description of  the three limiting geodesics at infinite $\lambda$. 
To that end, we  compute the solution set to  $|\rho_i|=1$ at infinite $\lambda$ and we find that the solutions lie on the curves represented on the left of Figure \ref{LimitCurves}. They meet at the branch points $z_b=e^{\pm i \frac{\pi}{3}}$.  
\noindent
We note that when we study the farthest geodesics at large but finite $\lambda$, we  set
\begin{equation}
|\rho_i|= e^{- \frac{\pi}{2 \lambda}} \, 
\end{equation}
to find the farthest geodesic. 
Moreover, recall from our approximation (\ref{rho1atorderlambdasquared}) to $\rho_1$ at a generic point at large $\lambda$ that there is a similar prefactor present in the expansion of $\rho_1$. Thus, the shape of the geodesics away from the branch point are only corrected at order $\lambda^{-2}$ at large $\lambda$. 
Attaining the regime of large $\lambda$ near the branch points numerically is non-trivial. We make use of the expression for $\rho_i$ at large $\lambda$ given in equation \eqref{AiryApproximation}. A numerical approximation to the geodesics at $\lambda=12$ is provided in the middle of Figure \ref{LimitCurves}. 
From these graphs as well as Figure \ref{HHHShortestGeodesics}, we see a picture emerging of the geodesics  around the singularities $z=0$ and $z=1$ meeting at the $\text{Re}(z)=1/2$ curve in the middle. The geodesic around $z=\infty$ snaps inwards towards the branch points. 

 Near the branch points, we expect the correction to the geodesics to be determined by the Airy function. The Airy function has an argument of order one for $\zeta$ of order $\lambda^{-\frac{2}{3}}$. Therefore, the correction to the path of the outermost geodesics near the branch point is of order $\lambda^{-\frac{2}{3}}$. The deviation at large $\lambda$ of the outermost geodesics from the limit curves  is therefore much stronger near the branch points as it is away from them. These qualitative expectations are borne out by the  features of our numerical graphs which show considerably stronger corrections to the geodesics near the branch point. 
We conclude that the large $\lambda$ approximation to the metric allows us to smoothly interpolate between the geodesics in Figure \ref{HHHShortestGeodesics} and the limit curves in Figure \ref{LimitCurves}.

\subsection{An Elliptic and Two Large Hyperbolic Monodromies}
For the hyperbolic-elliptic-hyperbolic solution, we study the limit where the elliptic singularity at $z=1$ is characterized by a finite parameter $\alpha_2$ and
we set the hyperbolic monodromies $\lambda_1=\lambda_3=\lambda$ equal and large.  The uni-valued metric  function
 around zero equals 
 \begin{align}
\rho_1 
&= e^{\frac{v_1}{\lambda}}\, z\, (1-z)^{-\frac{\alpha_2}{i\lambda}}\,  \left(\frac{{}_2 F_1 (\frac{1 - \alpha_2}{2}+i\lambda, \frac{1 - \alpha_2}{2} ,1 +i \lambda;z)}{{}_2 F_1 (\frac{1 + \alpha_2}{2} -i\lambda, \frac{1 + \alpha_2}{2} ,1 - i \lambda;z)}\right)^{\frac{1}{i\lambda}} \, .
\label{rho1HEH}
\end{align} 
We use the  Pfaff relation for the hypergeometric function 
\begin{align}
{}_2 F_1 (a, b, c, z) &= (1-z)^{-b}{}_2 F_1 (b, c-a, c, \frac{z}{z-1})~,
\end{align}
to rewrite the function $\rho_1$ in the form
\begin{align}
\label{rho1HEHasaratio}
    \rho_1 = e^{\frac{v_1}{\lambda}}\, z\, \left(\frac{{}_2 F_1 ( \frac{1 - \alpha_2}{2} , \frac{1 + \alpha_2}{2},1 +i \lambda;\frac{z}{z-1})}{{}_2 F_1 (\frac{1 + \alpha_2}{2} , \frac{1 - \alpha_2}{2} ,1 - i \lambda;\frac{z}{z-1})}\right)^{\frac{1}{i\lambda}}~,
\end{align}
where we have the overall coefficient
\be
e^{\frac{v_1}{\lambda}} = \left(\frac{\Gamma (1-i \lambda ) \sqrt{\Gamma \left(-\frac{\alpha _2}{2}+i \lambda +\frac{1}{2}\right)} \sqrt{\Gamma \left( \frac{\alpha _2}{2}+ i \lambda +\frac12\right)}}{\Gamma (1+i \lambda ) \sqrt{\Gamma \left(-\frac{\alpha _2}{2}-i \lambda +\frac{1}{2}\right)} \sqrt{\Gamma \left(\frac{\alpha _2}{2}- i \lambda +\frac12\right)}}\right)^{\frac{1}{i\lambda}}~.
\ee 
Again we consider the large $\lambda$ limit. The asymptotic expansion of the hypergeometric function in which the third of the parameters takes a large value is  given by \cite{Olver}:
\begin{align}
   \lim_{\kappa\rightarrow\infty}~ {}_2 F_1 (a,b,1+\kappa, z) = 1+ \frac{ab}{\kappa} z + \frac{a b z ((a+1) (b+1) z-2 )}{2 \kappa ^2} + \ldots~.
\end{align}
Near the hyperbolic singularity at $z=0$, we find a correction at second order to the local map:
\begin{equation}
\rho_1 = e^{-\frac{\pi}{2\lambda}} \left(
z+\frac{\left(\alpha _2^2-1\right) (z+1) z}{4 \lambda ^2 (z-1)}+ O(\lambda^{-4})
\right) \, .
\end{equation}
At second order in $\lambda^{-1}$ there is a divergence at $z=1$. Thus, the expansion is valid until $z$ approaches $1$ with a coordinate distance of order $\lambda^{-2}$. Near the elliptic singularity, we need a more refined large $\lambda$ approximation scheme. 

\subsubsection{Near the Elliptic Singularity}

Consider  the behavior of $\rho_1$ for large $\lambda$ but near the elliptic singularity at $z=1$. In effect this amounts to study the hypergeometric functions that appears in equation \eqref{rho1HEHasaratio}, 
\begin{equation}
{}_2 F_1 ( \frac{1 - \alpha_2}{2} , \frac{1 + \alpha_2}{2},1 \pm i \lambda; \frac{z}{z-1})~,
\end{equation}
for large argument, since $z$ is close to one. We use the results of \cite{KhawajaDaalhuis} to find a uniform approximation. The key  is an alternative integral representation for the  hypergeometric function:
\begin{align}
    {}_2F_1(a, b, b+\kappa, -\tilde{z}) =\frac{\Gamma(\kappa+b)}{\Gamma(\kappa)\Gamma(b)}\int_0^1 d\tau \frac{\tau^{b-1}(1-\tau)^{\kappa-1}}{(1+\tau \tilde{z})^a}~. 
\end{align}
In our set-up we have the parameter values
\begin{align}
\kappa &= \frac{1-\alpha_2}{2} \pm i \lambda \, , \quad a= \frac{1 - \alpha_2}{2}~, \quad \,  b = \frac{1 + \alpha_2}{2}~,\quad 
\tilde{z} = \frac{z}{1-z}~.
\end{align}
We can make a change of variables $\tau=1-e^{-t}$ to find the integral:
\begin{equation}
{}_2F_1(a,b,\kappa+b;-\tilde{z}) =
\frac{\Gamma(\kappa+b)}{\Gamma(\kappa)\Gamma(b)}\int_0^\infty dt \frac{t^{b-1}e^{-\kappa t}}{(1+\frac{t}{\zeta})^a} G(t)
\end{equation}
where
\begin{align}
G(t) &= \left( \frac{1-e^{-t}}{t} \right)^{b-1}
\left( \frac{1+ \frac{t}{\zeta}}{1+\tilde{z} -\tilde{z} e^{-t}} \right)^a~,\\
\zeta &= \log (1+\frac{1}{\tilde{z}})
= -\log z ~.
\end{align}
The saddle point approximation to this integral takes into account contributions from the endpoint $t=0$ of the contour as well as of the branch point at $t=-\zeta$ which coalesce for small $\zeta$ \cite{KhawajaDaalhuis}. 
This results in the following uniform expansion for the hypergeometric function  \cite{KhawajaDaalhuis}:
\begin{multline}
\frac{\Gamma(\kappa)}{\Gamma(\kappa+b)} {}_2 F_1 (a,b,\kappa; -\tilde{z}=\frac{z}{z-1}) = \zeta^b U(b,1 + \alpha_2,\kappa \zeta)
e^{-a \zeta} 
\nonumber \\
+ \zeta^{b+1} U(b,2+\alpha_2,\kappa \zeta) \frac{1-e^{-a \zeta} }{\zeta} + \dots
\end{multline}
The terms indicated by the ellipses are subleading terms in an expansion for large $\kappa$. 
In the small $\zeta$ fixed $\zeta \kappa$ expansion the first term is dominant. This allows us to truncate the  approximation to the hypergeometric function further:
\begin{align}
{}_2 F_1 (\frac{1-\alpha_2}{2},\frac{1+\alpha_2}{2},1+\kappa; 
\frac{z}{z-1}) & 
\approx \frac{\Gamma(\frac{1+\alpha_2}{2}+\kappa)}{\Gamma(\kappa)} 
\frac{e^{\frac{\zeta  \kappa }{2}} \zeta^{\frac12} \kappa^{-\frac{\alpha _2}{2}} K_{\frac{\alpha _2}{2}}\left(\frac{\zeta  \kappa }{2}\right)}{\sqrt{\pi }}
e^{-\frac{1-\alpha_2}{2} \zeta} ~. \nonumber
\end{align}
Substituting this into the expression for $\rho_1$, we find 
\begin{align}
\rho_1 &=
\left(\frac{1-\alpha_2-2i\lambda}{1-\alpha_2+2i\lambda}\right)^{\frac{\alpha_2}{2i\lambda}} 
\left(\frac{\Gamma(\frac12-\frac{\alpha_2}{2}-i\lambda)}{\Gamma(\frac12-\frac{\alpha_2}{2}+i\lambda) }
\frac{\Gamma(\frac12+\frac{\alpha_2}{2}+i\lambda)}{\Gamma(\frac12-\frac{\alpha_2}{2}+i\lambda)}\right)^{\frac{1}{2i\lambda}}
\left(
\frac{
K_{\frac{\alpha _2}{2}}\left(\frac{1}{4} \zeta  \left(1 -\alpha _2+ 2 i \lambda \right)\right)}
{K_{\frac{\alpha _2}{2}}\left(\frac{1}{4} \zeta  \left(1-\alpha _2-2 i \lambda \right)\right)} \right)^{\frac{1}{i \lambda}} \, .
\label{rho1HEHBesselK}
\end{align}
We can expand this expression for very large $\lambda$, large $\lambda \zeta$ and small $\zeta$ and we find agreement with the previous expansion for $\rho_1$ to order $-1$ in the coordinate distance to $z=1$. 

\begin{figure}[H]
\begin{center}
\includegraphics[width=6.5
cm]{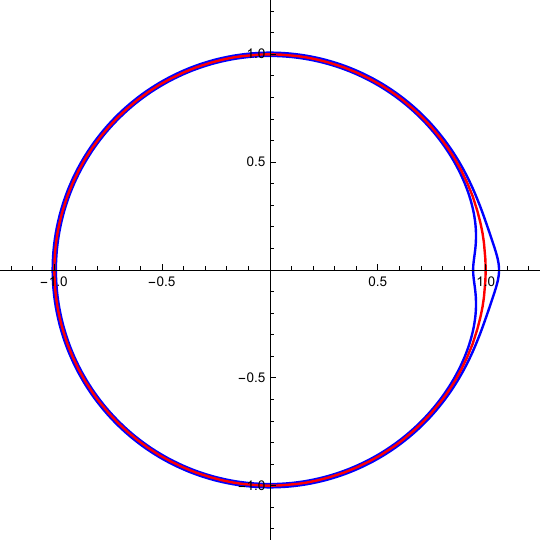}
\end{center}
\caption{For the value of $\lambda = 8$ and $\alpha_2 = 0.2$, we plot the farthest geodesics around $z=0$ and $z=\infty$ in blue and the limit curve in red. We note that the outermost geodesics deviate away from the elliptic singularity.}
\label{HEHshortestgeolimitcurve}
\end{figure}

\begin{figure}[H]
\begin{center}
\includegraphics[width=8cm]{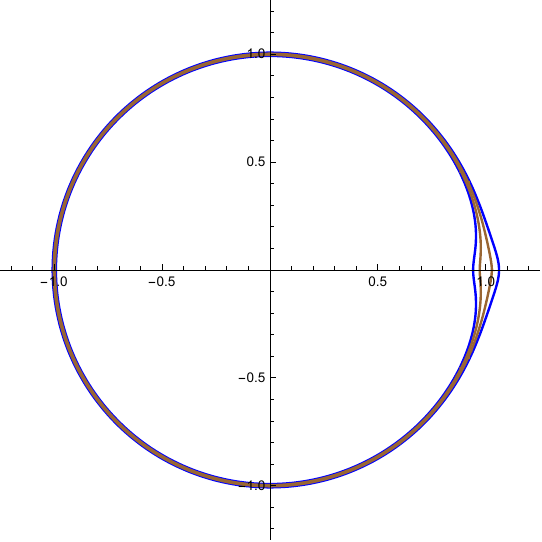}
\includegraphics[width=8cm]{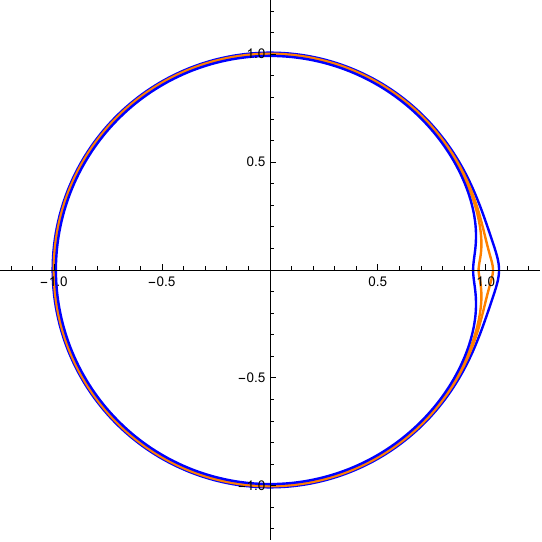}
\end{center}
\caption{On the left, we have the outermost geodesics drawn for the same value of $\lambda = 8$ and for $\alpha_2 = 0.2$ and $0.8$. The curve with the smaller value of $\alpha_2$ (in blue) and hence the larger deficit angle, leads to a bigger deviation away from the elliptic singularity. On the right we have geodesics drawn for the same value of $\alpha_2=0.2$, but for the values  $\lambda = 8$ and $\lambda=14$. The geodesic with the larger value of $\lambda$ (in orange) gets closer to the elliptic singularity.}
\label{plotswithdifflambdaandalpha}
\end{figure}

From the expression for $\rho$ in equation \eqref{rho1HEH}, we find that the limit of the farthest geodesic at infinite $\lambda$ is  the unit circle $|\rho_1|=1$. For the case of finite but large $\lambda$ and $\alpha_2 < 1$, one can again draw the farthest geodesics around the hyperbolic singularities at $z=0$ and $z=\infty$ using the expression for $\rho_1$ in equation \eqref{rho1HEHBesselK}. This is shown in Figure \ref{HEHshortestgeolimitcurve}. The limit curve lies between the farthest geodesics around zero  and infinity. 
In order to get an even better picture of the parameter space, we vary both $\alpha_2$ and $\lambda$ and see how the farthest geodesics vary; these results are shown in Figure \ref{plotswithdifflambdaandalpha}. 
In the figure, we have only drawn the farthest geodesics associated to each hyperbolic singularity. At this point it is worthwhile to recall our schematic Figure \ref{Fish}. As shown in that figure, the elliptic singularity is wedged in between the outermost geodesics around the two hyperbolic singularities. Furthermore, on either side of the farthest geodesic, one has a series of alternating singular curves and geodesics (radially inwards towards $z=0$ and radially outwards towards $z=\infty$).

\subsection{A Large Hyperbolic and Two Elliptic Monodromies}
Finally we study the case with two finite elliptic monodromies and one hyperbolic monodromy, which we take to be large. Once again, we consider the $\rho$ function defined around the hyperbolic singularity at $z=\infty$: 
\begin{equation}
   \rho_3= e^{\frac{v_3}{\lambda}} \, 
   z^{-1} (1- \frac{1}{z})^{\frac{i \alpha_2}{\lambda}} \left(
    \frac{{}_2{F}_1(\frac{1 + i \lambda - \alpha_2 + \alpha_1}{2}, \frac{1 + i \lambda - \alpha_2 - \alpha_1}{2} ,1 + i \lambda;\frac{1}{z}) }{{}_2{F}_1(\frac{1 - i \lambda + \alpha_2 -\alpha_1}{2}, \frac{1 - i \lambda + \alpha_2 + \alpha_1}{2} ,1 - i \lambda;\frac{1}{z}) } \right)^{\frac{1}{i \lambda}} ~,
\end{equation}
The asymptotics of the hypergeometric function that appears in $\rho_3$ have been studied in a number of references using steepest descent \cite{Watson} or a direct analysis of the hypergeometric differential equation \cite{Jones}. 
We follow the second reference  -- we  need the special case  $a=b$ in the formula (36) of \cite{Jones} --:
\begin{multline}
    {}_2 F_1(a+\kappa, a-c+1+\kappa, 1+2\kappa, \frac{2}{1-w}) \approx \frac{\Gamma(2\kappa+1)2^{\frac12-\kappa}(w-1)^{a+\kappa - \frac{c}{2}} (w+1)^{\frac{1}{2}(c-2a-1)}}{\Gamma(a+\kappa)\Gamma(\kappa-a+c)} \\
    \times \left(\frac{\sinh\zeta}{\zeta}\right)^{\frac12}\kappa^{c-1}\left[ \zeta K_{c-1}(\kappa\zeta)\left(1+O(\kappa^{-1})\right) - \zeta^2 K_c(\kappa\zeta) \left(\frac{B_0}{\kappa} +O(\kappa^{-2})\right)\right] \, .
\label{2F1asympforEEH}
\end{multline}
Here we have
\begin{align}
    w &= 1-2z~,\quad \zeta = \cosh^{-1}(w)~, \quad 
     \kappa = \pm \frac{i \lambda}{2}\, , \\
    a &= \frac12(1 \pm \alpha_1 \mp \alpha_2)~, \quad c= 1 \pm \alpha_1~. 
\end{align}
We moreover have the result:
 \begin{align}
 B_0 &= \frac{1}{2 \zeta} \left( (-\frac{1}{4} + (c-1)^2)(\frac{1}{\zeta}-\coth \zeta) + (\frac{1}{2} ((c-1)^2-1)- \frac{1}{2} ((2a-c)^2-1)) \tanh \frac{\zeta}{2} \right) \, . \nonumber
\end{align}
When we take the large $\lambda$ limit of the uniform asymptotics (with all other quantities fixed) then we obtain the function:
\begin{equation}
\rho_3 = -e^{- \frac{\pi}{2 \lambda}} 
e^{-\cosh ^{-1}(1-2 z)}
\left( 1+\frac{2 \pi  \left(\alpha _1+\alpha _2 \right) }{\lambda } +O(\lambda^{-2}) \right) \, .
\end{equation}
From this expression, we conclude that in the infinite $\lambda$ limit, the limit curve for  $|\rho_3|=e^{- \frac{\pi}{2 \lambda}}$ corresponds to the line segment between $z=0$ and $z=1$ on the real $z$-axis. For large but finite $\lambda$, one can use the expression for the hypergeometric function in terms of the first Bessel-K function in equation \eqref{2F1asympforEEH} and plot the outermost geodesic around the hyperbolic singularity at infinity, which corresponds to $|\rho_3| = e^{-\frac{\pi}{2\lambda}}$.  For fixed values of $\alpha_i$ and decreasing values of $\lambda$, these are shown in Figure \ref{Ellipse} below. 
\begin{figure}[H]
\begin{center}
\includegraphics[width=16cm]{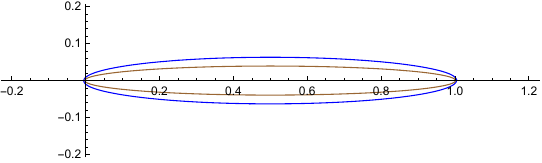}
\end{center}
\caption{The farthest geodesics are plotted for the case $\alpha_1=\alpha_2 = 0.2$. The limit curve, for $\lambda \rightarrow \infty$ is the straight line in red. The brown curve is for  $\lambda = 16$ and the blue curve for $\lambda = 10$. The outermost geodesics at smaller $\lambda$ values circumscribe those for the larger $\lambda$ values, and go around both elliptic singularities at $z=0$ and $z=1$.}
\label{Ellipse}
\end{figure}
\noindent
As one can see, these  enclose the limit curve and move outward as $\lambda$ is decreased. 
This figure should be compared to the schematic figure we drew in Figure \ref{EEHLogo}, in which the farthest geodesic can be seen to enclose the elliptic singularities at $z=0$ and $z=1$. Furthermore, for a particular value of $\lambda$, one again has an infinite sequence of alternating singular curves and geodesics going outward from the outermost geodesic.

\section{Two-Body Problems in Three-dimensional Gravity}
\label{GRin3D}
\label{3dGravity}
In the first part of the paper, we studied solutions to the Liouville equation with hyperbolic and elliptic monodromies. Due to the wide occurrence of the Liouville equation in problems in physics, these solutions will find  applications. As an illustration, in this second part of the paper we demonstrate the relevance of the solutions to two-body problems in three-dimensional gravity. We briefly review some relevant literature. 

The  static two  particle problem in three-dimensional space-time has a straightforward exact solution \cite{Deser:1983tn,Deser:1983nh}. The  classical solution for two moving massive particles in the case of zero cosmological constant  was described perturbatively in the velocities of the particle and exactly in the Newton constant times mass in 
\cite{Bellini:1995vm,Bellini:1995rw} and non-perturbatively in a first order formalism in \cite{Bellini:1995rz}. The solution was partially implicit. The static solution was given using solutions to the Liouville equation with elliptic monodromies in \cite{Welling:1997fw}.  
When the cosmological constant is negative, the static solution is also known exactly \cite{Deser:1983tn,Deser:1983nh} and can again be described as a solution to the Liouville equation \cite{Welling:1997fw}. In the case of moving particles, there is a perturbative approach to finding the metric \cite{Valtancoli:1999zg}. 

The second part of the paper is structured as follows. 
In section \ref{3dGR} we discuss the generic problem of solving Einstein's equations with sources and recall the features that render the problem exactly solvable in three dimensions.
We discuss a first example in which the Einstein equations are reduced to the Liouville equation which allows for the rapid deduction of an exact metric on a two-dimensional spatial slice as well as on the three-dimensional geometry \cite{Welling:1997fw}. We then adapt the solution generating technique to two particles moving in a space-time with negative cosmological constant. 
In section \ref{BlackHolesAndParticles} we discuss the three-dimensional geometries that explicitly describe either two relatively moving black holes, particles creating a black hole or a black hole absorbing a particle. We also discuss the link between the large hyperbolic monodromy metrics and very heavy black hole metrics.
We  conclude in section \ref{Conclusions}.

\subsection{Gravitating Point Sources in Three Dimensions}
\label{3dGR}
\label{Particles}

In this subsection we briefly review the Einstein field equations with point particle sources and why they are solvable in three dimensions \cite{Deser:1983tn,Deser:1983nh}.  We introduce the slicing of flat and anti-de Sitter space-times that reduces the Einstein equations to the Liouville equation \cite{Welling:1997fw}. We review to what extent two-particle metrics are known and generalize the metrics to include moving particles in $AdS_3$. 
\subsubsection{The Einstein Field Equations}
The Einstein field equations of general relativity are:
\begin{align}
R_{\mu \nu} - \frac{1}{2} g_{\mu \nu} R + \Lambda g_{\mu \nu} &= 8 \pi G T_{\mu \nu} \, .
\end{align}
They relate the Ricci tensor $R_{\mu \nu}$ of a metric $g_{\mu \nu}$ to the cosmological constant $\Lambda$ as well as to the matter energy-momentum tensor source $T_{\mu \nu}$. We have introduced a gravitational coupling $G$.  
The energy-momentum tensor $T_{\mu \nu}$ is necessarily covariantly conserved as a consequence of a Bianchi identity for the curvature tensor. 
In the following, we study the backreaction of point particle sources labeled by $i$ 
on the space-time metric $g_{\mu \nu}$. 
We parameterize a particle trajectory by a time $t$ and suppose they move on  trajectories $\xi_i^\mu(t)$.
 Provided the trajectories of the particles are geodesics, the energy-momentum tensor is covariantly conserved.  We suppose that we are dealing with geodesic particle trajectories throughout.

In three dimensions, there are no propagating gravitons and no local degrees of freedom in the metric. Therefore, outside the localized particle sources, the solution to general relativity is flat Minkowski space-time when the cosmological constant is zero and locally $AdS_3$ when the cosmological constant is negative \cite{Deser:1983tn,Deser:1983nh}. 
One method to include localized particle sources is to perform a coordinate transformation (on a locally flat or $AdS_3$ metric) which is singular on the particle trajectories. 

\subsubsection{The Metric in Flat Spacetime}

\label{TwoParticlesInFlatSpace}

\label{TwoMovingMassiveParticles}

We start from a flat space-time metric. We then perform a coordinate transformation that is ill-defined at the position of two moving point particle sources. The coordinate transformation is such that it  introduces the energy-momentum source associated to the  particles \cite{Deser:1983tn,Deser:1983nh}. For static sources, the problem of introducing two particle sources simultaneously was solved neatly in  \cite{Welling:1997fw}. In this reference, the choice of a coordinate system in which the Einstein equations reduce to the Liouville equation was stressed.  How to perform a similar operation for moving sources was described in \cite{Bellini:1995rz} in the first order formalism.  Our contribution here lies in combining these insights  and providing an efficient and simple solution to the moving two-particle metric that is  explicit. 

A solution to Einstein's field equations with point sources in three dimensions and with zero cosmological constant is locally given by the flat space metric:
\begin{align}
ds^2 &= -dT^2 + dX_i^2
\end{align}
everywhere outside the point sources. 
In order to apply the solution generating technique though, we prefer to describe the flat space-time with the metric  
\begin{equation}
ds^2 = -dt^2 + t^2 e^{\phi} dZ d \bar{Z} \, , \label{HyperbolicSliceMetric}
\end{equation}
where the vacuum corresponds to the choice of scale factor
\begin{equation}
e^{\phi} = \frac{4}{(1-Z \bar{Z})^2} \, ,
\end{equation}
and the coordinate sets are related by the transformations:
\begin{align}
X_1 &= t \sinh \theta \cos \phi \, , \qquad
X_2 
= t \sinh \theta \sin \phi \, , \qquad
T 
= t \cosh \theta \label{CoordinateTransfo1}
\end{align}
as well as
\begin{align}
Z &= \tanh \frac{\theta}{2} e^{i \phi} \, .
\label{CoordinateTransfo2}
\end{align}
The new coordinates cover the whole of the Minkowski space-time. The spacelike slice at constant time $t$ is a Poincar\'e disk with $SU(1,1) \approx SO(2,1)$ isometry group.  The space-time isometry subgroup $SO(2,1)$  is of particular interest to us and it is the isometry group of the hyperbolic spatial slices. The time coordinate $t$ is chosen such that $t^2=T^2-X^2-Y^2$ is invariant under rotations and boosts. This is a first reason to consider this particular coordinate system. 
A second reason to choose this coordinate system is that
a solution to the Liouville equation for the metric on the spatial slice gives a solution to the Einstein equations. 

The generic technique is as follows. We recall that a solution to the Liouville equation takes the form:
\begin{equation}
e^\phi = 4 \frac{\partial f(\zeta) \bar{\partial} \bar{f} (\bar{\zeta})}{(1-f \bar{f})^2} \, .
\end{equation}
From vacuum, we obtain the corresponding metric (\ref{HyperbolicSliceMetric}) through the coordinate transformation 
\begin{equation}
Z = f(\zeta) \, .
\end{equation}
We will allow solutions to the Liouville equation with singularities at three points such that we can incorporate two gravitating bodies and their asymptotics into our metric. Finally, we allow our particles to move on geodesics. Bringing these techniques together, we construct the solution as follows. 

To standardize the metric solution associated to generic geodesics $\xi_{1,2}(t)$, we perform a first coordinate transformation. 
We  introduce a coordinate $\zeta$ in which the particle sources are located at $\zeta=0$ and $\zeta=1$ at all times $t$:
 \begin{equation}
 \zeta = \frac{z-\xi_1(t)}{\xi_2(t)-\xi_1(t)}
 \, . \label{FractionalTransformation}
 \end{equation}
 The second step is to perform a holomorphic coordinate transformation $Z=f(\zeta)$
where the coordinate transformation has singular points at $\zeta=0,1$ as well as asymptotically at $\zeta=\infty$. These singular points have prescribed monodromy that code the particle masses $m_{1,2}$. The particles also have a  relative momentum which is captured by the total monodromy which we can measure at $\zeta=\infty$. We consider the case without intrinsic or orbital angular momentum -- the monodromies are then $SO(2,1)$ valued. 
The holomorphic function that succeeds in introducing precisely the desired monodromies is a ratio of solutions to the hypergeometric differential equation with three singularities with coefficients prescribed by the two masses $m_{1,2}$ and the total mass $m_3$ of the solution \cite{Welling:1997fw}.  It is the solution we discussed in section \ref{Liouville} in detail. The relation between the  angle parameters $\alpha_i$ and the masses $m_i$ is:
\begin{align}
\alpha_i  &= 1-4 Gm_i \, .
\end{align}
The solution is -- see equation (\ref{AllElliptic}) --
\begin{align}
f(\zeta) &= \zeta^{\alpha_1} \coth \frac{\delta}{2} \frac{\widetilde{F}(\frac{1}{2}
(1 + \alpha_3 + \alpha_1-\alpha_2),
\frac{1}{2} (1-\alpha_3 +\alpha_1-\alpha_2), 1 + \alpha_1; \zeta)}{\widetilde{F}(\frac{1}{2}
(1+\alpha_3-\alpha_1-\alpha_2),
\frac{1}{2} (1-\alpha_3-\alpha_1-\alpha_2), 1-\alpha_1; \zeta)} \, . \label{LiouvilleSolutionThreeElliptic}
\end{align}
The relation between the monodromy $\alpha_3$ at infinity and the relative velocity parameter $\delta$ was discussed in section \ref{Liouville}. 
It should be clear that by plugging in the coordinate transformation into the original flat metric (\ref{HyperbolicSliceMetric}), we obtain an entirely explicit metric solution by differentiation and  simple algebra. The solution is a generalization of the one in \cite{Welling:1997fw} since we allowed for the possibility of time-dependent geodesics $\xi_{1,2}(t)$. A limitation of the solution is that the angular momentum is zero. To incorporate angular momentum, we must allow for monodromies in the full $\mathfrak{iso}(2,1)$ gauge algebra. Those solutions would require a non-trivial  generalization of our techniques.

\subsubsection{The Metric in Anti-de Sitter Spacetime} 
\label{TwoParticlesInAdS3}
We can repeat these considerations for two moving massive point particles in the presence of a negative cosmological constant.
The two  particles are trapped inside the $AdS_3$ gravitational well and  interact with each other gravitationally.
Any spatial slice will contain two points $\xi_{i=1,2}$ which are the distinct positions of the two particles at a given time.
We consider the Einstein-Hilbert action with a negative cosmological constant $\Lambda=-2/l^2$ and with two massive particle sources. We often set $l=1$ for convenience. 
The locally $AdS_3$ space-time can be described as a patch of the hyperboloid:
\begin{align}
 l^2 &= U^2 +V^2-X^2 -Y^2 
 \end{align}
 and it has an $\mathfrak{so}(2,2)$ isometry algebra.
 We pick coordinates \cite{Welling:1997fw}:
\begin{align}
X &= l \cos \tau \sinh \theta \cos \phi
\, , \qquad
Y
= l \cos \tau \sinh \theta \sin \phi
\nonumber \\
U&= l \cos \tau \cosh \theta \, , \qquad \qquad
V
= l \sin \tau \label{AdS3PoincareDisks}
\end{align}
which can locally be gathered in the $SL(2,\mathbb{R})$ group element $g$:
\begin{align}
g &= \left( \begin{array}{cc}
V+Y &  U+X \\
-U+X    & V-Y
\end{array} \right)
\, .
\end{align}
We can view the spatial isometry group as a diagonal subgroup of the $SO(2,2)$ group acting left and right on the group element $g$. We have that $U^2-X^2-Y^2=\cos^2 \tau$ and $V=\sin \tau$ are constant on a constant time slice. The entry $V$ is half the trace of the $SO(2,1)$ group element $g$. Under a diagonal $g \rightarrow h g h^{-1}$ action the trace of the group element $g$ and therefore also the time $\tau$ will be  invariant. Thus, the  diagonal $SO(2,1)$ group action has again been rendered geometric on a Poincar\'e disk. 
 
 We  perform a further coordinate transformation
 \begin{equation}
 Z = \tanh \frac{\theta}{2} e^{i \phi}   \label{ZPoincareDisk}
 \end{equation}
 such that  the  locally $AdS_3$ solution has a metric
 \begin{equation}
 ds^2 = -d\tau^2 + \cos^2 \tau \frac{4 dZ d \bar{Z}}{(1-Z\bar{Z})^2} .
\end{equation}
Our preference is for a complexified coordinate system with a Poincar\'e   spatial slice with isometry group $SO(2,1)$ in order to apply the singular coordinate transformation technique and to exploit the solutions to the Liouville equation once more \cite{Welling:1997fw}. 
Indeed, if we generalize the metric to:
\begin{equation}
 ds^2 = -d\tau^2 + \cos^2 \tau e^{\phi} dZ d \bar{Z} .
 \label{AdSLiouville}
\end{equation}
then the scalar field $\phi$ again satisfies the Liouville equation.
 At $\tau=\pi/2+ \pi \mathbb{Z}$, the overall scale of the spatial section becomes zero. The coordinate system ceases to be valid. 
%
Locally, the monodromy problem around the singularities is identical to the one in flat space-time. The singular coordinate transformation to introduce the desired holonomies is therefore also identical and we take once more the transformation (\ref{LiouvilleSolutionThreeElliptic}), based on the elliptic Liouville solution (\ref{AllElliptic}),
\begin{equation}
Z = f(\zeta(z,t)) \, .
\end{equation}
After the double coordinate transformation, we find an exact description of the massive two moving particle metric in $AdS_3$.
With the fractional  coordinate transformation (\ref{FractionalTransformation}), this metric becomes a dynamical generalization of the static two-particle metric described in \cite{Welling:1997fw}.

\subsection{Black Hole and Particle Metrics}
\label{BlackHolesAndParticles}
In section \ref{Particles} we discussed how a solution of Einstein's equations in three dimensions with particle sources can be obtained from solutions to the Liouville equation with elliptic monodromies. In section \ref{Liouville} we discussed solutions to the Liouville equation based on auxiliary doublets with elliptic and hyperbolic monodromies. In this section we exploit the latter to obtain explicit three-dimensional metrics of black holes and particles. The resulting geometries have been discussed in various contexts in terms of hyperbolic geometries with identifications \cite{Brill:1995jv,Steif:1995pq}. We believe it is useful to write down explicit metrics for these solutions. Since the background reasoning on how the Liouville solutions become solutions to three-dimensional general relativity is identical to the one for particles, described in section \ref{Particles}, we only highlight new features. 

\subsubsection{One Black Hole and Two Asymptotics}
For starters, we consider the  simpler geometry of a single hyperbolic monodromy around a point on the Poincar\'e disk. Automatically, we also have a hyperbolic monodromy at infinity. This  gives rise to two asymptotic regions as we propagate further away or near the singularity. We want to square this description of the geometry \cite{Brill:1995jv,Steif:1995pq} with the canonical description of the BTZ black hole in three-dimensional space-time in terms of Schwarzschild type coordinates. It is not hard to do this using the description of the global geometry \cite{Banados:1992gq}. 

Recall that we wrote the three-dimensional metric with spatial slices that have a conformal factor given by a Liouville solution.
When we study the part of the group manifold covered by our coordinate system, we find that it covers the patch with  $|V| \le l$. We do cover the whole of the $Z$ plane. However, the analogue of our $z$ coordinate only covers a part of the $Z$ Poincar\'e  disk -- see \cite{Welling:1997fw}. This part is  drawn in Figure 1 of  \cite{Brill:1995jv} or Figure 2 of \cite{Welling:1997fw}. The relation to BTZ coordinates can be established by following the coordinate transformation from $z$ to $Z$ to the group coordinates, and then from the group elements to the BTZ coordinates \cite{Banados:1992gq}. The map can be worked out and is complicated. To summarize, the region that is drawn in \cite{Brill:1995jv,Welling:1997fw} is a region between two boundaries, with a horizon in the middle. This region lies between given values for the modulus of $z$ -- see \cite{Welling:1997fw}. 

If we connect this to the description of the solutions to Liouville theory, we find the following picture. Firstly, we have the singularity at $z=0$. It is surrounded by circles on which the metric degenerates. We have annuli 
in which the metric is well-defined and each annulus contains a horizon. Boundaries alternate with horizons continuously infinitely close and infinitely far from the hyperbolic singularity. One can choose 
a region 
as a fundamental region for the spatial slice. It has two asymptotes. Such a fundamental region is what is considered in the references \cite{Brill:1995jv,Welling:1997fw}.

\subsubsection{Two Black Holes and Three Asymptotes}

\label{BigBlackHoles}

We turn to the explicit metric solutions modeling two black holes in the interior and a third asymptotic region which is also shielded by a horizon. These geometries were abstractly described in \cite{Brill:1995jv,Steif:1995pq}.
We have learned from \cite{Hadasz:2003he,Firat:2021ukc} that the geometry we obtain is an extension that glues an infinite number of hyperbolas to the outside legs -- see Figure \ref{PantsWithDiabolos}. We can  amputate the extensions and reduce the space-time to that of \cite{Brill:1995jv}. The spatial slice we describe is time-symmetric. The geodesics are the locations of horizons \cite{Steif:1995pq}. The latter are minimal area or constant mean curvature subspaces inside the time-symmetric spatial slice \cite{Steif:1995pq}. Therefore, our description of geodesics inside the spatial slice becomes a snapshot of the horizon at the turning point in time.  
The conclusion is that the three-dimensional metrics (\ref{AdSLiouville}) with the Liouville  input  \cite{Hadasz:2003he,Firat:2021ukc} gives the explicit metrics for the solution that \cite{Brill:1995jv} describes in an abstract manner.
Our analysis of geodesics becomes an analysis of horizons. The large hyperbolic monodromy limit becomes a description of the geometry in the limit of very massive black holes. We can reformulate all of our statements in sections \ref{Liouville} and  \ref{LargeHyperbolicMonodromy} in a gravitational language. This informs us about solutions to three-dimensional general relativity. It also inspires generalizations of the previous analysis. As an example, note that we took all hyperbolic monodromies equal and tending to infinity. Obviously, we can generalize this to black holes with a certain mass ratio as well as different relative velocities. The resulting saddle point problems are known to be interesting and intricate \cite{ParisII,CFU,Watson,Jones,KhawajaDaalhuis}, and related to string field theory vertices \cite{Firat:2021ukc}. We leave them for further study.

\subsubsection{Mixed Particle/Black Hole Two-Body Metrics}
Finally, using the Liouville solutions with mixed  hyperbolic and elliptic monodromies, we have in hand the explicit metric for a particle in the presence of a black hole or two particles with a relative momentum such that the asymptotic monodromy becomes hyperbolic. These space-times have been described qualitatively in \cite{Matschull:1998rv,Steif:1995pq}
We have provided explicit metrics. They are found by combining the equation (\ref{AdSLiouville}) for the locally $AdS_3$ metric in terms of the Liouville field with the hyperbolic-elliptic-hyperbolic Liouville solution or the elliptic-elliptic-hyperbolic solution based on the ratio of auxiliary fields in equations (\ref{HEHSolution}) and (\ref{EEHSolution}) respectively. 
 The elliptic-elliptic-hyperbolic set-up is where a  black hole is formed by colliding two point particles at high speed. Moreover, the large black hole mass analysis of section \ref{LargeHyperbolicMonodromy} gives a simpler explicit expression for the special functions appearing in the metric and allows for the numerical analysis of the snapshot of the black hole horizon as in the Figures in section \ref{LargeHyperbolicMonodromy}. For brevity,  we illustrate this with but one example. Figure \ref{plotswithdifflambdaandalpha} can be read as an analysis of a  snapshot of the horizon geometry on a time-symmetric slice when we vary the mass of the black hole or the mass of the particle in the geometry. 

Once more, the mathematical physics of the Liouville equation finds direct application in three-dimensional gravity and the analysis of explicit physical limits (e.g. the limit of large black hole masses) leads to saddle point integration as well as interesting special function theory. Vice versa, questions in three-dimensional gravity will feed back into which properties of the geometries to  study further.

\section{Conclusions}
\label{Conclusions}
We studied classical solutions to the Liouville equation based on an auxiliary set of fields with elliptic or hyperbolic monodromy around three points on a compactified complex plane. The detailed description of the solutions requires a careful analysis of $SL(2,\mathbb{R})$ or $SU(1,1)$ valued monodromies. We described the generic properties of the resulting two-dimensional metrics. These differ depending on the mix of monodromies. We  moreover proposed the generic picture that may arise from analyzing the set of outermost geodesics around hyperbolic singularities. 

We then turned to determining the solutions in the limit of large hyperbolic monodromies. This gives rise to interesting saddle point approximations to integral representations for the hypergeometric function ${}_2F_1$, including an example in which Riemann introduced the saddle point method. We exploited more recent contributions to the vast literature in this field  to describe a few example metrics in detail, using simpler special functions. These analyses allowed us  to provide analytic approximations to the description of the most outward geodesic in a given homotopy class associated to a hyperbolic singularity. 

The Liouville equation surfaces in many contexts. One is the classical limit of the ubiquitous two-dimensional Liouville conformal field theory. In this setting, our new solutions  describe the classical limit of mixed elliptic-hyperbolic correlation functions. We focused on another application in which the Liouville equation describes the geometry of a two-dimensional slice of a solution to three-dimensional gravity. The outermost geodesic we analyzed becomes a snapshot of a black hole horizon. The large monodromy limit is a large mass black hole limit. The  Liouville solution provides an explicit metric and the mixed cases correspond to  particle-black hole geometries. We demonstrated that, by pushing the Liouville analysis, we learn more about the gravity solutions. 

There are plenty  of open problems in this seemingly simple domain. For instance we assumed that the analysis of the outermost geodesic around a given hyperbolic singularity is largely independent of the nature of the other singularities. A proof of this claim remains an outstanding problem.
%
When we considered the large monodromy limit, we considered the  case in which all the hyperbolic monodromies were scaled to be large in an identical fashion. The analysis of the large monodromy limit with relative weights  remains to be carried out. This is interesting from a mathematical point of view \cite{ParisII,CFU,Watson,Jones,KhawajaDaalhuis} as well as for analyzing large mass black hole metrics. 

From a physics perspective, there are many further interesting questions to explore. One would like to have a coordinate frame in which one can follow the evolution of horizons until they merge. Adding angular momentum to the two particle metric while maintaining the exactness of the solution in a coordinate system that is continuous is also an open task. 
Many  problems persist at the intersection of Liouville theory, special function theory and three-dimensional relativity. We believe they continue to be worth studying since they have an elementary and universal nature.

\section*{Acknowledgements}

We are grateful to Dharmesh Jain and Alok Laddha for helpful discussions and to Eleonora Dell'Aquila for help with the graphics. SA would like to thank the Universit\`a di Torino, Italy for their hospitality during the completion of this work.

\appendix

\section{The Details of the Classical Liouville Solutions}
In this Appendix, we gather formulas that describe the classical Liouville solutions in the bulk of the paper in more detail. 

\subsection{All Hyperbolic Monodromies}
\label{AppendixHyperbolicMonodromies}
We consider the case where all three monodromies are hyperbolic.  
We begin with an auxiliary doublet $\psi^{\pm}$ with a diagonal monodromy near $z=0$: 
\be 
\begin{pmatrix}
    \psi_1^+\\
    \psi_1^-
\end{pmatrix} \rightarrow 
\begin{pmatrix}
    -e^{-\pi \lambda_1} & 0 \\
    0 & - e^{\pi \lambda_1}
    \end{pmatrix} 
    \begin{pmatrix}
    \psi_1^+\\
    \psi_1^-
\end{pmatrix} ~.
\ee
We will denote this monodromy matrix as ${M_1}^1$. 
We similarly assume a diagonal hyperbolic monodromy $M_2^2$ around $z=1$ for the doublet $\psi_2^{\pm}$:
\be 
\begin{pmatrix}
    \psi_2^+\\
    \psi_2^-
\end{pmatrix} \rightarrow 
\begin{pmatrix}
    -e^{-\pi \lambda_2} & 0 \\
    0 & - e^{\pi \lambda_2}
    \end{pmatrix} 
    \begin{pmatrix}
    \psi_2^+\\
    \psi_2^-
\end{pmatrix} ~,
\ee
and a diagonal monodromy $M_3^3$ around $z=\infty$ for the doublet $\psi_3^{\pm}$: 
\be 
\begin{pmatrix}
    \psi_3^+\\
    \psi_3^-
\end{pmatrix} \rightarrow 
\begin{pmatrix}
    -e^{-\pi \lambda_3} & 0 \\
    0 & - e^{\pi \lambda_3}
    \end{pmatrix} 
    \begin{pmatrix}
    \psi_3^+\\
    \psi_3^-
\end{pmatrix} ~.
\ee
The three sets of wavefunctions are given by
\begin{align}
\psi_1^\pm &= \frac{e^{\pm \frac{i v_1}{2}}}{\sqrt{i \lambda_1}} z^{\frac{1\pm i \lambda_1}{2}}
(1-z)^{\frac{1 \mp i \lambda_2}{2}} 
{}_2 F_1 \left(\frac{1 \pm i \lambda_1 \mp i \lambda_2 \pm i \lambda_3}{2}, \frac{1 \pm i \lambda_1 \mp i \lambda_2 \mp i\lambda_3}{2} ,1 \pm i \lambda_1;z\right)
\nonumber \\
\psi_2^\pm &= i \frac{e^{\pm \frac{i v_2}{2}}}{\sqrt{i \lambda_2}} (1-z)^{\frac{1\pm i \lambda_2}{2}}
z^{\frac{1 \mp i \lambda_1}{2}} 
{}_2 F_1 (\frac{1 \pm i \lambda_2 \mp i \lambda_1 \pm i\lambda_3}{2}, \frac{1 \pm i \lambda_2 \mp i \lambda_1 \mp i\lambda_3}{2} ,1 \pm i \lambda_2;1-z)
\nonumber \\
\psi_3^\pm &= (iz) \frac{e^{\pm \frac{i v_3}{2}}}{\sqrt{i\lambda_3}} (\frac{1}{z})^{\frac{1\pm i \lambda_3}{2}}
(1-\frac{1}{z})^{\frac{1 \mp i \lambda_2}{2}} \nonumber\\
&\hspace{2cm}{}_2 F_1 (\frac{1 \pm i\lambda_3 \mp i \lambda_2 \pm i \lambda_1}{2}, \frac{1 \pm i\lambda_3 \mp i \lambda_2 \mp i \lambda_1}{2} ,1 \pm i\lambda_3;\frac{1}{z}) \, .
\end{align}
We have to demand that the monodromy is SL$(2,\mathbb{R})$ valued  around all punctures.  For this purpose, we can use the connection formulae for the hypergeometric functions in order to relate the doublet around $z=0$ to those with diagonal monodromy at  the punctures $z=1$ and $z=\infty$. 
Using the connection matrices ${S_1}^2$ and ${S_1}^3$, one can derive the monodromy of the basis $\{\psi_1^+, \psi_1^- \}$ around $z_2=1$ and $z_3=\infty$ by composing the matrices: 
\begin{align} 
{M_1}^2 &=  {S_1}^2 {M_2}^2 ({S_1}^2)^{-1} \cr
&= \left(
\begin{array}{cc}
 -\cosh (\gamma ) \sinh \left(\pi  \lambda _2\right)-\cosh \left(\pi  \lambda _2\right) & -\sinh (\gamma ) \sinh \left(\pi  \lambda _2\right) \\
 \sinh (\gamma ) \sinh \left(\pi  \lambda _2\right) & \cosh (\gamma ) \sinh \left(\pi  \lambda _2\right)-\cosh \left(\pi  \lambda _2\right) 
\end{array}
\right)  . 
\end{align} 
In order to write the monodromy matrix ${M_1}^2$ in this compact form, we have substituted the expression obtained for the complex phase $e^{i v_1}$ in \eqref{expv1} and also used the definition \eqref{HHHRelation} of $\gamma$. Similarly we can define 
\begin{align} 
{M_1}^3 &=  {S_1}^3 {M_3}^3 ({S_1}^3)^{-1} \\
&=\left(
\begin{array}{cc}
 e^{\pi  \lambda _1} \left(\cosh \left(\pi  \lambda _2\right)-\cosh (\gamma ) \sinh \left(\pi  \lambda _2\right)\right) & -e^{\pi  \lambda _1} \sinh (\gamma ) \sinh \left(\pi  \lambda _2\right) \\
 e^{-\pi  \lambda _1} \sinh (\gamma ) \sinh \left(\pi  \lambda _2\right) & e^{-\pi  \lambda _1} \left(\cosh (\gamma ) \sinh \left(\pi  \lambda _2\right)+\cosh \left(\pi  \lambda _2\right)\right) 
\end{array}
\right)  . \nonumber
\end{align}
We have verified that these matrices satisfy the consistency condition:
\begin{equation}
{M_1}^1 \cdot {M_1}^3\cdot {M_1}^2 = \mathbb{I}\, .
\label{MultiplyToOne}
\end{equation}

\subsection{Mixed Monodromies}

Let us consider a mixed case in which there are hyperbolic monodromies at $z=0$ and $z=\infty$ and an elliptic monodromy at $z=1$. We start out with the  ansatz for the wavefunctions:
\begin{align}
\psi_1^\pm &= \frac{e^{\pm \frac{i v_1}{2}}}{\sqrt{i \lambda_1}} z^{\frac{1\pm i \lambda_1}{2}}
(1-z)^{\frac{1 \mp \alpha_2}{2}} 
{}_2 F_1 \left(\frac{1 \pm i \lambda_1 \mp \alpha_2 \pm i \lambda_3}{2}, \frac{1 \pm i \lambda_1 \mp \alpha_2 \mp i\lambda_3}{2} ,1 \pm i \lambda_1;z\right)
\nonumber \\
\xi_2^\pm &= i \frac{e^{\pm \frac{i v_2}{2}}}{\sqrt{\alpha_2}} (1-z)^{\frac{1\pm \alpha_2}{2}}
z^{\frac{1 \mp i \lambda_1}{2}} 
{}_2 F_1 (\frac{1 \pm \alpha_2 \mp i \lambda_1 \pm i\lambda_3}{2}, \frac{1 \pm \alpha_2 \mp i \lambda_1 \mp i\lambda_3}{2} ,1 \pm \alpha_2;1-z)
\nonumber \\
\psi_3^\pm &= (iz) \frac{e^{\pm \frac{i v_3}{2}}}{\sqrt{i\lambda_3}} (\frac{1}{z})^{\frac{1\pm i \lambda_3}{2}}
(1-\frac{1}{z})^{\frac{1 \mp \alpha_2}{2}} \nonumber\\
&\hspace{2cm}{}_2 F_1 (\frac{1 \pm i\lambda_3 \mp \alpha_2 \pm i \lambda_1}{2}, \frac{1 \pm i\lambda_3 \mp \alpha_2 \mp i \lambda_1}{2} ,1 \pm i\lambda_3;\frac{1}{z}) \, .
\end{align}
As in the case of three hyperbolic monodromies, we apply the connection formula that relates the solutions near $z=0$ to the solutions with diagonal monodromy near $z=1$
\begin{align}
    \psi^+_1 &=-\frac{e^{\frac{\pi i}{4}} \sqrt{\alpha_2} e^{\frac{1}{2} i \left(v_1-v_2\right)} \Gamma \left(1+i \lambda _1\right) \Gamma \left(-\alpha_2\right) }{\sqrt{\lambda _1} \Gamma \left(\frac{1}{2} \left(i \lambda _1-i \lambda _3-\alpha_2+1\right)\right) \Gamma \left(\frac{1}{2} \left(-\alpha_2+i \left(\lambda _1+\lambda _3-i\right)\right)\right)}\xi_2^+ \cr
    &-\frac{e^{\frac{\pi i}{4}} \sqrt{\alpha_2} e^{\frac{1}{2} i \left(v_1+v_2\right)} \Gamma \left(1+i \lambda _1\right) \Gamma \left(\alpha_2\right) }{\sqrt{\lambda _1} \Gamma \left(\frac{1}{2} \left(i \lambda _1-i \lambda _3+m_2+1\right)\right) \Gamma \left(\frac{1}{2} \left(m_2+i \left(\lambda _1+\lambda _3-i\right)\right)\right)}
     \xi^-_{2}
\end{align}
\begin{align}
    \psi_1^- &= -\frac{e^{\frac{\pi i}{4}} \sqrt{\alpha _2} e^{-\frac{1}{2} i \left(v_1+v_2\right)} \Gamma \left(-\alpha _2\right) \Gamma \left(1-i \lambda _1\right) }{\sqrt{\lambda _1} \Gamma \left(\frac{1}{2} \left(-\alpha _2-i \lambda _1+i \lambda _3+1\right)\right) \Gamma \left(\frac{1}{2} \left(-\alpha _2-i \left(\lambda _1+\lambda _3+i\right)\right)\right)}\xi^+_{2} \cr
    &-\frac{e^{\frac{\pi i}{4}} \sqrt{\alpha _2} e^{-\frac{1}{2} i \left(v_1-v_2\right)} \Gamma \left(\alpha _2\right) \Gamma \left(1-i \lambda _1\right) }{\sqrt{\lambda _1} \Gamma \left(\frac{1}{2} \left(\alpha _2-i \lambda _1+i \lambda _3+1\right)\right) \Gamma \left(\frac{1}{2} \left(\alpha _2-i \left(\lambda _1+\lambda _3+i\right)\right)\right)}\xi^-_{2} \, .
\end{align}
The coefficients of the wavefunctions define the entries of the connection matrix ${S_1}^2$ that links the solutions near $z=0$ to those at $z=1$. The $\psi_2^{\pm}$ doublet is such that it has a diagonal elliptic monodromy at $z=1$ and the monodromy 
\be 
M_2^2 = \begin{pmatrix}
    -e^{i\pi \alpha_2} & 0 \\
    0 & -e^{-i \pi \alpha_2}
\end{pmatrix}~
\ee 
is an element of the group SU$(1,1)$.
Thus the monodromy of the solutions $\psi_1^{\pm}$ around $z=1$ is obtained by composing the monodromies $M_1^2 = {S_1}^2 M_2^2 ({S_1}^2)^{-1}$. In order to fix the undetermined coefficient $v_1$ in this case, we demand that $M_1^2$ be real and that it belongs to SL$(2,\mathbb{R})$. 
Substituting this into the monodromy matrix $M_1^2$ and using the variable $\eta$ introduced in equation \eqref{etadefn}
we obtain the compact expression
\be 
M_1^2=
\left(
\begin{array}{cc}
 -\csc (\eta ) \sin \left(\pi  \alpha _2+\eta \right) & \sin \left(\pi  \alpha _2\right) \csc (\eta ) \\
- \sin \left(\pi  \alpha _2\right) \csc (\eta ) & -\csc (\eta ) \sin \left(\eta -\pi  \alpha _2\right) \\
\end{array}
\right) \, .
\ee 
One can similarly use the connection formula to write the solutions near $z=0$ in terms of those near $z=\infty$. We have
\begin{align}
    \psi_1^+ &= -\frac{i \sqrt{\lambda _3} \Gamma \left(1+i \lambda _1\right) \Gamma \left(-i \lambda _3\right) e^{\frac{1}{2} \left(\pi  \left(\lambda _1+\lambda _3\right)+i v_1-i v_3\right)}}{\sqrt{\lambda _1} \Gamma \left(\frac{1}{2} \left(i \lambda _1-i \lambda _3-m_2+1\right)\right) \Gamma \left(\frac{1}{2} \left(i \lambda _1-i \lambda _3+m_2+1\right)\right)}\psi_3^+\cr
    &-\frac{i \sqrt{\lambda _3} \Gamma \left(1+i \lambda _1\right) \Gamma \left(i \lambda _3\right) e^{\frac{1}{2} \left(\pi  \left(\lambda _1-\lambda _3\right)+i v_1+i v_3\right)}}{\sqrt{\lambda _1} \Gamma \left(\frac{1}{2} \left(-m_2+i \left(\lambda _1+\lambda _3-i\right)\right)\right) \Gamma \left(\frac{1}{2} \left(m_2+i \left(\lambda _1+\lambda _3-i\right)\right)\right)}\psi_3^- \\
    \psi_1^- &= -\frac{i \sqrt{\lambda _3} \Gamma \left(1-i \lambda _1\right) \Gamma \left(-i \lambda _3\right)  e^{\frac{1}{2} \left(-\pi  \lambda _1+\pi  \lambda _3-i \left(v_1+v_3\right)\right)}}{\sqrt{\lambda _1} \Gamma \left(\frac{1}{2} \left(-m_2-i \left(\lambda _1+\lambda _3+i\right)\right)\right) \Gamma \left(\frac{1}{2} \left(m_2-i \left(\lambda _1+\lambda _3+i\right)\right)\right)}\psi_3^+\cr
    &-\frac{i \sqrt{\lambda _3} \Gamma \left(1-i \lambda _1\right) \Gamma \left(i \lambda _3\right) e^{\frac{1}{2} \left(-\pi  \left(\lambda _1+\lambda _3\right)-i v_1+i v_3\right)}}{\sqrt{\lambda _1} \Gamma \left(\frac{1}{2} \left(-i \lambda _1+i \lambda _3-m_2+1\right)\right) \Gamma \left(\frac{1}{2} \left(-i \lambda _1+i \lambda _3+m_2+1\right)\right)}\psi_3^- ~.
\end{align}
The coefficients of the wavefunctions $\psi_3^{\pm}$ define the connection matrix ${S_1}^{3}$. Now the monodromy of the solutions $\psi_3^{\pm}$ around $z=\infty$ is diagonal, and given by
\be 
M_3^3 = \begin{pmatrix}
    -e^{-\pi\lambda_3} & 0 \\
    0 & - e^{\pi \lambda_3}
\end{pmatrix}~.
\ee 
Thus the monodromy of the solutions $\psi_1^{\pm}$ around $z=\infty$ is obtained by composition with the connection matrix
\begin{align} 
M_1^3 &= {S_1}^{3}~M_3^3~({S_1}^{3})^{-1} \cr 
& = \left(
\begin{array}{cc}
 e^{\pi  \lambda _1} \csc (\eta ) \sin \left(\eta -\pi  \alpha _2\right) & e^{\pi  \lambda _1} \sin \left(\pi  \alpha _2\right) \csc (\eta ) \\
 -e^{-\pi  \lambda _1} \sin \left(\pi  \alpha _2\right) \csc (\eta ) & e^{-\pi  \lambda _1} \csc (\eta ) \sin \left(\pi  \alpha _2+\eta \right) \\
\end{array}
\right)~.
\end{align} 
The monodromy matrices again satisfy the consistency condition (\ref{MultiplyToOne}). 

A similar analysis can be carried out for the case with two elliptic and one hyperbolic monodromies.
The results have been summarized in the main text.

\bibliographystyle{JHEP}

\end{document}